\documentclass[aps,prl,twocolumn,a4paper,floatfix,10pt]{revtex4-1}
\usepackage{graphicx,amsmath,amssymb,amsfonts,dsfont,enumitem,MnSymbol}

\usepackage{cancel}
\newcommand{\ket}[1]{\ensuremath{|#1\rangle}}
\usepackage{etoolbox}
\patchcmd{\thebibliography}{\section*{\refname}}{}{}{}
\DeclareUnicodeCharacter{0308}{~} 

\usepackage{svg}
\UseRawInputEncoding
\usepackage{soul}
\usepackage{amsmath}
\usepackage{multirow}
\usepackage[normalem]{ulem}
\usepackage{hyperref}

\usepackage{blindtext}

\usepackage{color}
\usepackage[caption=false]{subfig}

\usepackage{float}
\RequirePackage{hyperref}
\usepackage{subfiles}

\usepackage{amsmath}

\begin{document}
\title{Quantum Information Perspective on Many-Body Dispersive Forces}
\author{Christopher Willby${}^{1,2}$}
\author{Martin Kiffner${}^{3}$}
\author{Joseph Tindall${}^{4}$}
\author{Jason Crain${}^{5,1}$}
\author{Dieter Jaksch${}^{2,1}$}
%
\affiliation{${}^1$Clarendon Laboratory, University of Oxford, Parks Road, Oxford OX1
3PU, United Kingdom}
\affiliation{${}^2$Institut f{\"u}r Quantenphysik, Universit{\"a}t Hamburg, 22761 Hamburg, Germany}
\affiliation{${}^3$PlanQC GmbH, Lichtenbergstr. 8, 85748 Garching, Germany}
\affiliation{${}^4$Centre for Computational Quantum Physics, Flatiron Research Institute, New York, New York 10010 USA}
\affiliation{${}^5$IBM Research Europe, The Hartree Centre STFC Laboratory, Sci-Tech Daresbury, Warrington WA4 4AD, United Kingdom}

\begin{abstract}
Despite its ubiquity, the quantum many-body properties of dispersion remain poorly understood. Here, we investigate the entanglement distribution in assemblies of quantum Drude oscillators, minimal models for dispersion-bound systems. We establish an analytic relationship between entanglement and correlation energy and show how entanglement monogamy determines whether many-body corrections to the pair potential are attractive, repulsive, or zero. These findings, demonstrated in trimers and extended lattices, apply in more general chemical environments where dispersion coexists with other cohesive forces.

\end{abstract}

\maketitle

\textit{Introduction.-} Any pair of neutral atoms or molecules experiences an attraction due to the formation of transient electric dipoles. Though weaker than other forms of binding, this so-called dispersion force is the only inter-molecular interaction present in all materials: It  spans a wide range of length-scales and its cumulative effects can exert decisive influence over physical, chemical, and biological processes ~\cite{stone2013theory,hirschfelder2009intermolecular,london1930theorie,hermann2017first,grimme2011density,distasio2014many,pastorczak2017intricacies,von2004optimization,french2010long,dzyaloshinskii1961general}. These include stability of condensed phases in noble gases ~\cite{stillinger1985computer,axilrod1951triple,jansen1964stability}, hydrocarbons \cite{rösel2017london}, molecular crystals \cite{kronik2014understanding,wagner2015london} as well as non-covalent binding of drugs  to proteins ~\cite{distasio2012collective,stohr2019quantum,tkatchenko2011unraveling} and even the ability of geckos to climb up flat surfaces ~\cite{autumn2002evidence,singla2021direct}.

In many cases, dispersion is approximated in material simulations as a pair-wise interaction attenuated by a $1/R^6$ distance dependence. However, failures of the pairwise approximation and the importance of many-body corrections are now widely acknowledged in numerous contexts including supramolecular chemistry~\cite{ambrosetti2014hard},  clusters ~\cite{tkatchenko2012accurate},  low dimensional systems including layers, chains and wires ~\cite{misquitta2014anomalous}, nanomaterials~\cite{shtogun2010many,hermann2017nanoscale,deringer2016many}, inorganic \cite{deringer2016many} and alkaline earth compounds~\cite{kim2016failure}, as well as polymorphism in  molecular crystals~\cite{marom2013many,distasio2014many,reilly2014role,reilly2013seamless}. In rare gas condensates,  many-body effects can account for nearly 10\% of the cohesive energy~\cite{rosciszewski2000ab} and in layered systems, their contributions can be much larger~\cite{maaravi2017interlayer}.

The elementary picture of dispersion interactions between two particles is that of ``flickering" dipole moments, arising from correlated zero-point fluctuations of the molecular or atomic charge density.  However, such a model has no immediate or systematic generalisation to the many-body case. Consequently, it is not known a-priori, when or why the pairwise approximation fails, when perturbative corrections will be sufficient, or whether or under what circumstances the net contributions from many-body effects will be attractive or repulsive. Ultimately, the residual interactions governing both pairwise and many-body dispersion are quantum phenomena with  no classical analogue. 

Correlated vacuum fluctuations give rise to entangled ground states, where the state of each particle cannot be described independently of the others.
Going from the pairwise to the many-body case however, will incur certain restrictions on the allowed amount of entanglement two particles may share in a multipartite system \cite{terhal2004entanglement,adesso2006entanglement,horodecki2009quantum}. 
These restrictions, known as the monogamy of entanglement \cite{coffman2000distributed,osborne2006general,hiroshima2007monogamy,wolf2003entanglement,coffman2000distributed,osborne2006general,pawlowski2010security,lloyd2014unitarity,seevinck2010monogamy,koashi2004monogamy,hiroshima2007monogamy,o2001entangled,ferraro2007monogamy}, are expressed by inequalities that give upper bounds for the pairwise entanglement. Within the confines of the monogamy inequalities the amount of shareable  entanglement increases with the dimensionality of the local Hilbert space of the particles ~\cite{adesso2007coexistence,adesso2007entanglement,adesso2006entanglement}. For infinite dimensional Gaussian states, there is thus a rich interplay between the enhancement of pairwise entanglement in the multipartite setting and its limitations due to monogamy restrictions~\cite{ferraro2007monogamy}.

Despite being considered a key property of entanglement, the ramifications of the monogamy property on chemical binding, are still largely unexplored. Moreover, finding the relationship between binding energy and entanglement measures remains mostly uncharted, with some work relating correlation energy to entropy measures~\cite{cioslowski2024constraints,collins1993entropy}.

Here, we derive a bound on the pairwise and many-body dispersive energy contributions in terms of the reduced tangle, defined via a monogamous entanglement metric for Gaussian states. The tightness of this bound is demonstrated in both a trimer and arbitrary lattice structures.
For the trimer, we partition the parameter space into a monogamous regime, where monogamy constraints suppress pairwise entanglement and a promiscuous regime, where the multipartite interaction allows for enhanced pairwise entanglement.
The boundary separating these regimes closely aligns with the transition between repulsive and attractive many-body energy contributions, along which  pairwise additivity remains valid.
We further show that the relationship between monogamy constraints and many-body energy contributions persists in larger systems. Since the trimer captures key features of many-body dispersion in lattices, our findings provide a quantum information perspective on many-body energetics.   

Finally, we consider how our methods and conclusions extend to realistic chemical environments. Our findings apply both to scenarios where dispersion is the sole cohesive force and to more general cases where it coexists with other forms of bonding. This generality arises because we focus on quantum fluctuations around an equilibrium configuration. While other interactions may shift the equilibrium positions, the fluctuations around these points persist. As such, our entanglement-based characterization of dispersion remains applicable across a wide range of chemical systems, from dispersion-dominated noble-gas crystals to dispersion-influenced molecular complexes.

\textit{Model.-} We model many-body dispersion forces at long range in molecular systems as assemblies of dipole coupled quantum Drude oscillators (QDOs)~ \cite{cipcigan2019electronic,jones2013electronically,vaccarelli2021quantum,bade1957drude,wang2001drude,jones2013quantum,tkatchenko2012accurate,donchev2006many,cipcigan2016electronic,london1937general,sparnaay1959additivity,hirschfelder1964molecular,cao1992many,sadhukhan2016quantum}.  The QDO model  has been demonstrated to capture the dominant physics of non-reactive condensed matter systems, and outperforms competitor models in terms of accuracy and computational efficiency, in many instances \cite{sadhukhan2016quantum,bryenton2023many,xu2020many,khabibrakhmanov2023universal,goger2023optimized}. In particular, the dipole-order QDO Hamiltonian is employed to amend many-body dispersive corrections into density functionals,  via the so-called many-body dispersion (MBD) method ~\cite{tkatchenko2012accurate,ambrosetti2014long,distasio2014many,reilly2014role,reilly2013seamless,ambrosetti2014hard},  the accuracy of which is well established in the literature. See the supplemental material (SM) for a greater discussion of the applicability of the dipole QDO model, through MBD.

In the QDO model a particle, molecule or molecular fragment is treated as a three dimensional quantum harmonic oscillator (given by three quantum harmonic oscillators or modes per QDO), with mass $m_{\mu}$, frequency $\omega_{\mu}$, nuclear charge $+q_{\mu}$ and Drude quasi-particle of charge $-q_{\mu}$. 
The instantaneous dipole moment $q_{\mu}\boldsymbol{\hat{r}}_{\mu}$ is proportional to the displacement, $\boldsymbol{\hat{r}_{\mu}}=(\hat{r}^{x}_{\mu},\hat{r}^{y}_{\mu},\hat{r}^{z}_{\mu})$ of the QDO from its equilibrium, determined by the center position of the particle.  The corresponding momentum vector is given by $\boldsymbol{\hat{p}}_{\mu}=(\hat{p}^{x}_{\mu},\hat{p}^{y}_{\mu},\hat{p}^{z}_{\mu})$.

Dipolar interactions between the QDOs arise from the Coulomb potential and are described by a $3N\times 3N$ coupling matrix $\mathcal{T}$. Individual $3 \times 3$ blocks of $\mathcal{T}$ describe the interaction between QDO $\mu$ and $\xi$, given by
\begin{equation}\label{dipoledipole}
e^{T}_{\mu}\mathcal{T}e_{\xi}=
\frac{1}{R^{3}_{\mu \xi}}\left(\mathds{1}_{3}- \frac{3 \boldsymbol{R}_{\mu \xi} \otimes \boldsymbol{R}_{\mu \xi}}{R^{2}_{\mu \xi}}\right)  \quad{} \text{for} \quad{}  \mu \neq \xi,
\end{equation}
and $e^{T}_{\mu}\mathcal{T}e_{\mu}=0_{3}$, where $0_{3}$ is the $3 \times 3$ zero matrix. Here $\otimes$ denotes the outer product,
$e^{T}_{\mu}$ is a $3 \times 3N$ matrix, e.g $ e^{T}_{1}=[\mathds{1}_{3},0_{3},..0_{3}],e^{T}_{2}=[0_{3},\mathds{1}_{3},..0_{3}]$ etc and
$\boldsymbol{R}_{\mu \xi}=(x_{\mu}-x_{\xi},y_{\mu}-y_{\xi},z_{\mu}-z_{\xi})$, where $(x_{\mu},y_{\mu},z_{\mu})$ is the position of the QDO in Cartesian space.

For simplicity we here consider the case where the QDO parameters are all identical given by $(q,m,\omega)$, for the general case see the SM. We define operators $\boldsymbol{\hat{\chi}}_{\mu}=\boldsymbol{\hat{r}}_{\mu}\sqrt{\omega m/\hbar}$, $\boldsymbol{\hat{\mathcal{P}}}_{\mu}=\boldsymbol{\hat{p}}_{\mu}/\sqrt{\hbar m \omega}$. Then by introducing the polarizability  $\alpha=q^{2}/(m\omega^{2}4\pi \epsilon_{0})$, the dipole order QDO Hamiltonian in units of $\hbar\omega$ is
\begin{equation}\label{sameQDOham}
    \hat{H}=\frac{1}{2}\left(\sum^{3N}_{i=1}\mathcal{\hat{P}}_{i}^{2}+\sum^{3N}_{i=1}\hat{\chi}_{i}^{2}\right)+\alpha\sum^{3N}_{i>j}\hat{\chi}_{i}\mathcal{T}_{ij}\hat{\chi}_{j}.
\end{equation}
We use the indices $i,j$ to label the individual harmonic oscillators or modes. 

\textit{Dispersive Binding.-} The dispersive binding energy is the difference between the ground state energy of the non-interacting and interacting QDOs. The dispersive binding energy, inclusive of the many-body contributions, is written as $ E=3N/2-(\sum^{3N}_{i=1}\sqrt{\lambda_{i}})/2$,
where $\lambda_{i}$ are the normal eigenvalues of the $3N \times 3N$ potential matrix $V$, with $V_{ii}=1$ and $V_{ij}=\alpha \mathcal{T}_{ij}, \: \forall i \not =j$. The expression for the dispersive binding energy $E$ follows from the adiabatic connection fluctuation dissipation theorem and can be found in \cite{langreth1977exchange,gunnarsson1976exchange}. The binding energy $E$ is always positive (see SM for details). In order for the energy to be real, the matrix $V$ must be positive definite. This enforces the constraint $\alpha t_{i} > -1 \: \ \forall i$, where $t_{i}$ are the eigenvalues of the real symmetric matrix dipole coupling $\mathcal{T}$.


The dispersive binding energy is often treated as a perturbative correction to the non-interacting ground state \cite{london1930theorie,jeziorski1994perturbation,kim2006van,langbein1971microscopic}, leading to the familiar pair-potential. This is derived by a second order Taylor expansion of $\sqrt{\lambda_{i}}$ in small parameters $\alpha t_{i}$. This is possible when $\alpha ||\mathcal{T}||_{2}<1$, equivalent to $\mathrm{max}(\alpha|t_{1}|,..,\alpha|t_{3N}|)<1$.
We will assume $\alpha ||\mathcal{T}||_{2}<1$ holds throughout, explicitly stating instances when this is not the case.
As $\sum^{3N}_{i=1}t^{k}_{i}=\mathrm{tr}(\mathcal{T}^{k})$ and  $\mathcal{T}$ is traceless, the full power series expansion of $E$ in terms of $\alpha t_{i}$ can be written as follows,
\begin{equation}\label{Eseries}
    \begin{split}
       E=\sum^{\infty}_{k=2} \delta_{k}, \quad{} \delta_{k}=\frac{(-1)^{k}(2k-3)!!}{2^{k+1}k!}\alpha^{k}\mathrm{tr}(\mathcal{T}^{k}).
    \end{split}
\end{equation}

The pairwise potential $\delta_2$, aptly named as it can be written additively over all pairs in the system, gives the leading order correction to the non-interacting energy  ~\cite{renne1967microscopic,donchev2006many}, 
\begin{equation}\label{pairwise}
\delta_{2}=\frac{\alpha^{2}}{16}\mathrm{tr}(\mathcal{T}^{2})=\frac{3\alpha^{2}}{4} \sum^{N}_{\substack{\mu>\xi}} R_{\mu \xi}^{-6}>0.
\end{equation}
The many-body (MB) correction to the binding energy is thus given by $\delta_{\rm MB}=E-\delta_{2}$ and can be repulsive or attractive. The so called Axilrod-Teller (AT)  triple dipole approximation is given by $\delta_{3}$ \cite{axilrod1943interaction,cao1992many}.  The MB effect can thus be written in terms of the higher order corrections to the pairwise potential,
$\delta_{MB}=\sum^{\infty}_{k=3}\delta_{k}$. 
\begin{figure*}
 \centering
    \includegraphics{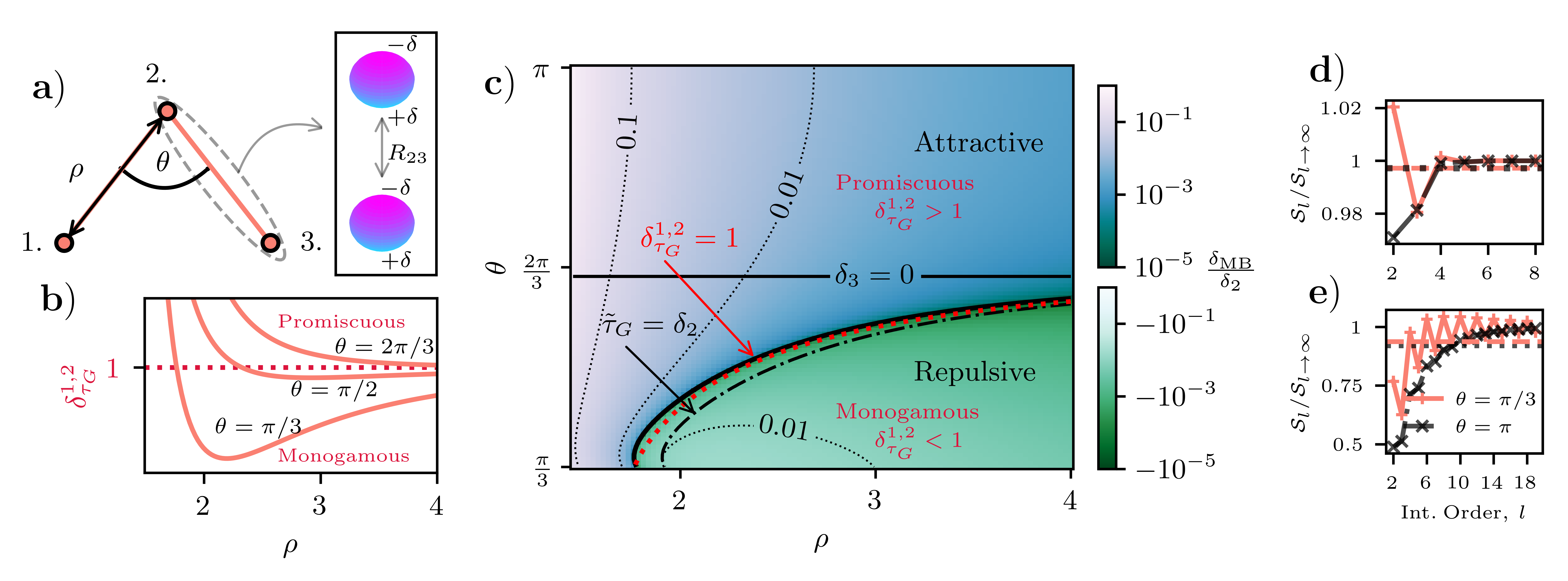}
     \caption{$\mathbf{a)}$ Depiction of the dipole interacting trimer setup. The correlated charge density fluctuations are shown by the instantaneous dipoles, denoted as $\pm \delta$. The coordinates of the three QDOs $1,2$ and $3$ are $(0,0,0)$,  $(R_{12}\sin{[\theta/2]},R_{12}\cos{[\theta/2]},0)$ and $(2R_{12}\sin{[\theta/2]},0,0)$, where $\theta \in \{\pi/3,\pi\}$. The trimer depends on a dimensionless nearest neighbor separation $\rho$ and geometry, $\theta$. $\mathbf{b)}$ Behavior of the nearest neighbor EDI,  $\delta^{1,2}_{\tau_{G}}$,  at different angles, with the dotted red line separating promiscuous and monogamous regions. 
     $\mathbf{c)}$ Above the
     red dotted line the sum of the pairwise tangle shared between the nearest neighbor QDOs behaves promiscuously, whereas below it behaves monogamously. The monogamy/promiscuity boundary overlaps the black line where MB energy corrections vanish and the pairwise additive approximation holds; above which MB corrections are attractive and below which are repulsive; both shown as heat-maps, plotted on a log scale. In $\mathbf{d)-e)}$, where $\rho=2.60$ and $\rho=1.55$ respectively, the crosses show $S_{l}$, where $S_{l}=\sum^{l}_{k=1}(k-1)\delta_{k}$, normalized by $S_{l \rightarrow \infty}=\sum^{\infty}_{k=1}(k-1)\delta_{k}$. The latter is computable from the CM matrix elements, (see SM for calculation details). The dashed and dotted horizontal lines show the normalized value of the reduced tangle, i.e $\tilde{\tau}_{G}/S_{l\rightarrow \infty}$ for the respective $\rho$ and $\theta$ values (dashed line for $\pi/3$ and dotted line for $\pi$).}\label{trimerheatmap} 
\end{figure*}

\textit{Gaussian States.-} The perturbative approach, can be readily compared to the exact solution of the quadratic QDO Hamiltonian, which has a Gaussian ground state, that can be calculated efficiently~\cite{weedbrook2012gaussian,schuch2006quantum,wolf2004entanglement,serafini2017quantum,derezinski2017bosonic,colpa1978diagonalization,van1980note,maldonado1993bogoliubov,plenio2005entropy,cramer2006entanglement,wolf2004entanglement,giedke2003entanglement,giedke2001entanglement,adesso2014continuous}. Gaussian ground states are completely characterized by a vector of single element correlation functions and a correlation matrix (CM) of two point correlation functions \cite{schuch2006quantum,manuceau1968quasi,holevo1971generalized}. As the equilibrium displacement of the QDOs is set to zero, we can describe the properties of the ground state, solely by the CM.
The ground state CM can be written in terms of the potential matrix $V$ \cite{audenaert2002entanglement,schuch2006quantum,cramer2006entanglement},
\begin{equation}\label{CM}
\gamma_{0}=V^{-1/2} \oplus V^{1/2},
\end{equation}
where the matrix square root is given by $V^{1/2}=O\mathrm{diag}(\sqrt{\lambda_{1}},..,\sqrt{\lambda_{3N}})O^{T}$, and $O$ is the orthogonal matrix that diagonalizes $V$, where 
$(V^{-1/2})_{ij}=\langle \hat{\chi}_{i}\hat{\chi}_{j} \rangle$ and 
$(V^{1/2})_{ij}=\langle \hat{\mathcal{P}}_{i}\hat{\mathcal{P}}_{j}\rangle$.

\textit{The Tangle.-} We quantify the entanglement present in the ground state by the Gaussian tangle which is an entanglement measure, i.e vanishing for separable states and non-increasing under local operations and classical communication (LOCC), with a verifiable monogamy property \cite{hiroshima2007monogamy,adesso2007entanglement,amico2008entanglement}. 
The tangle quantifies the entanglement in a bipartition between mode $i$ and the other $3N-1$ modes in the system, 
\begin{equation}\label{tangledensitymatrix}
\tau_{G}\left(i\right)=\frac{1}{4}\left(||\hat{\rho}^{T_{i}}||_{1}-1\right)^{2},
\end{equation}
where $\hat{\rho}$ is a \textit{pure} density matrix of a Gaussian state and $\hat{\rho}^{T_{i}}$ is the partial transpose with respect to the $i$th mode. From the CM framework, $||\hat{\rho}^{T_{i}}||_{1}=\sqrt{\langle \hat{\chi}_{i}^{2} \rangle \langle \hat{P}_{i}^{2} \rangle}+\sqrt{\langle \hat{\chi}_{i}^{2} \rangle \langle \hat{P}_{i}^{2} \rangle-1}$ \cite{hiroshima2007monogamy,vidal2002computable}.

The pairwise tangle, shared between modes $i$ and $j$ is denoted by $\tau_{G}(i:j)$. This is computable for a mixed state in terms of the convex roof construction. See the SM for a definition of $\tau_{G}(i:j)$ and Refs.  \cite{wolf2004gaussian,adesso2005gaussian} for further details.  
 The monogamy inequality imposes a constraint on the sum of pairwise tangles, $\tau_{G}(i) \geq \sum^{3N}_{j=1}\tau_{G}(i:j), \forall j \not = i$. The monogamy constraint arises from upper bounds on each pairwise tangle,  $f(-\langle \hat{\chi}_{i}\hat{\chi}_{j} \rangle \langle \hat{\mathcal{P}}_{i}\hat{\mathcal{P}}_{j}\rangle) \geq \tau_{G}(i:j)$   $\forall i \not= j$, where $ f(x)=(\sqrt{x}+\sqrt{x+1}-1)^{2}/4$  for $x>0$ and $f(x)=0$, given $x<0$, \cite{hiroshima2007monogamy}. If $\langle \hat{\chi}_{i}\hat{\chi}_{j} \rangle \langle \hat{\mathcal{P}}_{i}\hat{\mathcal{P}}_{j}\rangle>0$, then $\tau_{G}(i:j)=0$, due to the positive partial transpose (PPT) criterion~\cite{simon2000peres,adesso2004extremal}. 

\textit{Results.-} We study the entanglement and energy in terms of a dimensionless parameter $\rho=R/\alpha^{1/3}$, with $R=\mathrm{min}_{\mu,\xi}R_{\mu \xi}\forall \mu \neq  \xi$. In addition to $\rho$, the entanglement and energy depend on the geometry of the fixed equilibrium positions of the $N$ QDOs.
By using the simple cubic lattice as a reference frame, the authors in \cite{donchev2006many}, estimated the region $2.3 \leq \rho \leq 3$, to be an interval of primary importance, for solids and liquids.
\begin{figure*}
    \centering
 \includegraphics{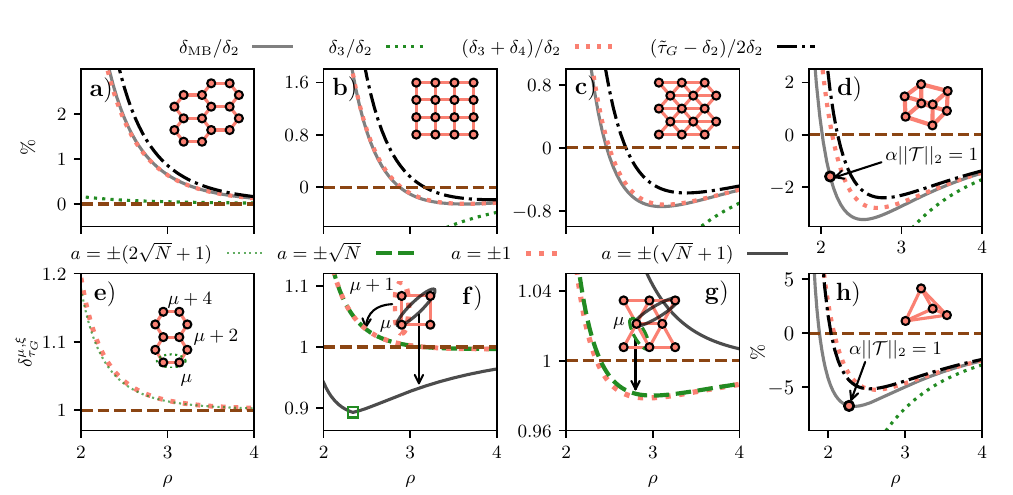}
    \caption{ Behavior of the MB effects and  entanglement on various 2D and 3D lattices. In $\mathbf{a)-d)}$ and $\mathbf{h)}$ the solid line shows the MB effects, the small green dotted line the AT correction, the large orange dotted line shows four-body corrections added to the AT potential, the dashed dotted line shows the deviation between the sum of the reduced tangle of each of the $3N$ modes and the pairwise bonding energy, $(\tilde{\tau}_{G}-\delta_{2})/2$, with the factor of $1/2$ included as $(\tilde{\tau}_{G}-\delta_{2})/2 \rightarrow \delta_{3}$ for $\rho \rightarrow \infty$. All the curves have been normalized by the pairwise bond energy and have been plotted as a function of the dimensionless nearest neighbor separation $\rho$. The following $2$D lattices are shown; $\mathbf{a)}$ $N=47 \times 47 \times 2$ honeycomb lattice $\mathbf{b)}$ $N=71 \times 71$ square lattice $\mathbf{c)}$ $N=71 \times 71$ triangular lattice. The $3$D lattices are $d)$ $N=19 \times 19 \times 19$ cubic lattice and $\mathbf{h)}$ $N=4\times 12\times 12\times 12$ pyrochlore lattice. All lattices calculations have been preformed with open boundary conditions. 
    The EDI between $\mu$ and $\xi=\mu+a$ on the $2$D lattices is shown in $\mathbf{e)-g)}$. The dashed horizontal line shows the monogamy/promiscuity boundary for the respective lattices, with the monogamous region lying below and the promiscuous region above.
    The ordering of the QDOs on the 2D lattices is shown in the insets in $\mathbf{e)-g)}$, where $\mu=2256$ in $\mathbf{e)}$, and $\mu=2520$ in $\mathbf{f)-g)}$. The open green square in $\mathbf{f)}$ shows where the PPT criterion ensures $\tau_{G}(3\mu:3\xi)=0$.}
        \label{lattices}
\end{figure*}

In the ground state of the two-mode Hamiltonian,  
$\hat{H}_{ij}=(1/2)(\hat{\chi}_{i}+\hat{\chi}_{j}+\hat{\mathcal{P}}_{i}+\hat{\mathcal{P}}_{j})+\alpha\mathcal{T}_{ij}\hat{\chi}_{i}\hat{\chi}_{j} $, 
the inequalities on the pairwise tangle, described in the previous section, saturate. We shall label the pairwise tangle computed from the ground state of the \textit{bipartite} Hamiltonian $\hat{H}_{ij}$ as 
$\tau_G^{\mathrm{ref}}(i:j)=(\sqrt{\epsilon_{+}/\epsilon_{-}}-1)^{2}/4$,  
with $\epsilon_{\pm}=\sqrt{1\pm \alpha|\mathcal{T}_{ij}|}$ (see SM for derivation).
The reference tangle $\tau_G^{\mathrm{ref}}(i:j)$ captures the entanglement shared between modes $i$ and $j$, due to the dimensionless dipole bond $\alpha \mathcal{T}_{ij}$, as if all other modes were screened away. Conversely, in the ground state of the \textit{multipartite} QDO Hamiltonian in Eq. \eqref{sameQDOham}, the upper bound on the pairwise tangle is written as
$\tau_G^{\mathrm{sys}}(i:j)=f(-\langle \hat{\chi}_{i}\hat{\chi}_{j} \rangle \langle \hat{\mathcal{P}}_{i}\hat{\mathcal{P}}_{j}\rangle)$,  
accounting for the indirect contributions and collective correlations present in the full ground state system, involving all $3N$ modes.

In order to numerically demonstrate the geometric and $\rho$ dependence of the monogamy constraints on the pairwise tangle, contained within a given two QDO subsystem, we define the two QDO entanglement distribution index (EDI), 
\begin{equation}\label{rcf}
\delta^{\mu,\xi}_{\tau_{G}}=\frac{\sum_{i,j}\tau^{\mathrm{sys}}_{G}(i:j)}{\sum_{i,j}\tau^{\mathrm{ref}}_{G}(i:j)}
    \quad{} \forall i \in \mu,  j \in \xi.
    \end{equation}
The EDI quantifies how monogamy constraints shape the entanglement distribution in a two QDO subsystem. For sufficiently weak interactions only the lowest two levels of each QDO can be occupied. The monogamy constraint can then limit the pairwise tangle leading to $\delta_{\tau_{G}}^{\mu,\xi}<1$, reflecting the \textit{monogamous} behavior of entanglement in qubit systems. For stronger interactions higher lying QDO levels may be occupied thus increasing the size of the accessible local Hilbert space and the overall amount of shareable entanglement. In this \textit{promiscuous} regime $\delta_{\tau_{G}}^{\mu,\xi}>1$ (for details see the simple three mode examples in the SM). 

We consider three QDOs (trimer); see Fig.~\ref{trimerheatmap} \(a)\) for a 
depiction of the setup. We focus on the EDI between QDOs 1 and 2 in the trimer. 
By symmetry, this is identical to the subsystem of QDOs 2 and 3, 
highlighted by the two QDOs encircled by the dashed line 
within the three-QDO trimer in Fig.~\ref{trimerheatmap} \(a)\). Figure  \ref{trimerheatmap} $b)$ shows that for the most acute trimer geometries, the monogamy constraint restricts nearest-neighbor pairwise entanglement sharing at weak interactions, but this restriction is lifted as the interaction strength increases.

Figure \ref{trimerheatmap} $c)$ shows that both the geometric and $\rho$ dependence of the entanglement distribution directly correlates with the MB potential in the trimer.
The parameter space of the 
trimer is partitioned into its promiscuous and monogamous regions, 
separated by the dotted red line. 
The boundary between these two regions approximates the locus of points 
that separate the attractive and repulsive MB energies. For any 
\(\rho \geq 2.3\), the deviation in \(\theta\) between the two boundaries is 
less than \(1.2\%\) of the \(\theta\) value where \(\delta_{\rm MB} = 0\). 
This correspondence becomes exact in the limit 
\(\rho \to \infty\), where the boundary between the promiscuous and 
monogamous regimes asymptotically overlaps with the parameter space 
where the AT potential vanishes, i.e., \(\delta_{3} = 0\), indicated by 
the horizontal black line.

Next we substantiate our numerical findings by  analytically relating measures of the ground state entanglement to the binding energy. We use the definition of the tangle to  introduce the entanglement measure $\tilde{\tau}_{G}(i)=g(\tau_{G}(i))$, defined here as the reduced tangle, where 
$g(x)=x(1+\sqrt{x})^{2}/(1+2\sqrt{x})^{2}$. The function $g(x)$ is  monotonically increasing  for all positive $x$,  ensuring the reduced tangle is an entanglement monotone under LOCC. We derive an inequality capturing the relation and dependence of pairwise and many-body contributions to the binding energy and the distribution of the reduced tangle as follows  (see SM for proofs),
\begin{equation}
 \begin{split}\label{tangletildebounds}
  \tilde{\tau}_{G}=\sum^{3N}_{i=1}\tilde{\tau}_{G}(i) \leq \sum^{\infty}_{k=2} (k-1)\delta_{k}.
\end{split}
\end{equation}


Figure \ref{trimerheatmap} $d)-e)$ shows the convergence to the bound in Eq. \eqref{tangletildebounds} for different couplings strengths. 
Taking $\delta_{2}$ away from both sides of Eq. \eqref{tangletildebounds}, shows that $\tilde{\tau}_{G}-\delta_{2} \leq \sum^{\infty}_{k=3}(k-1)\delta_{k}$. 
Comparing this bound with the definition of the MB energy and the form of $E$ in Eq. \eqref{Eseries} explains the qualitative correspondence between the dashed-dotted black line in
Fig. \ref{trimerheatmap} $c)$ and the sign of $\delta_{\rm MB}$, where the former divides the parameter space, into the regions where $\tilde{\tau}_{G}< \delta_{2}$
(below) and the regions where $\tilde{\tau}_{G} > \delta_{2}$.

The Gaussian state framework can be directly extended to arbitrary assemblies such as layered systems and 3D lattices, which we consider next. The features of the trimer, shown in Fig. \ref{trimerheatmap} $c)$, can be used to explain the sign of the MB contributions in the more complex lattices, shown in Fig. \ref{lattices}. At weak coupling, the sign of the MB effects are determined by the AT potential in the lattices. The AT potential can be written as a sum over triplets (trimers) of QDOs, (see SM). The honeycomb lattice has a positive and negligible AT potential, following from where $\delta_{3}=0$ in Fig. \ref{trimerheatmap} $c)$, where the primitive translation vectors of the honeycomb lattice form an angle of $2\pi/3$. As  the honeycomb lattice has a negligible AT potential, the MB effects in this lattice geometry only become visible through higher order terms. In Fig. \ref{lattices} $b)-d)$ and $h)$, where the AT potential is repulsive, a sign change occurs at strong coupling and the MB effects become attractive. The sign change also occurs in Fig. \ref{trimerheatmap} $c)$, as a function of $\rho$, for all geometries with repulsive AT corrections. Figure \ref{lattices} $a)-d)$ and $h)$, further show that $(\tilde{\tau}_{G}-\delta_{2})/2 \approx \delta_{\rm MB}$, on all the lattices at $\rho=4$, indicated by the overlap of the solid and dashed dotted lines.

Figure~\ref{lattices} $e)-g)$ displays the EDI for nearest- and next-nearest-neighbor QDO pairs 
near the center of honeycomb, square, and triangular lattices. In the honeycomb lattice, promiscuous entanglement sharing persists among its three nearest neighbors 
across all \(\rho\) values, reflecting its exclusively attractive MB potential. In contrast, the square and triangular lattices exhibit a transition from monogamous to promiscuous 
entanglement sharing as \(\rho\) decreases and the interaction strength increases. The deviation between the \(\rho\) value 
with average nearest neighbor EDI value equal to 1 and 
the \(\rho\) value at which the MB potential vanishes is \(2.98\%\) for the triangular lattice and \(9.01\%\) for 
the square lattice, relative to the \(\rho\) value where \(\delta_{\rm MB} = 0\). 
This suggests that the boundary between monogamous and promiscuous regions provides a 
good approximation for the \(\rho\) value at which the pairwise additive approximation holds on $2$D lattices
(see SM for further evidence of EDI tracking the sign of the MB potential in the linear and zigzag chains).

 \textit{Conclusion.-}  We show that the many-body (MB) dispersive interaction energy is controlled,
 via an inequality, by an entanglement monotone defined in terms of a monogamous entanglement measure for Gaussian states. This constitutes, to the best of our knowledge, the first analytically rigorous relation between an entanglement measure and the significant contribution to the chemical correlation energy, due to dispersion.

We characterize 
the MB energetics through the monogamous or promiscuous behavior of entanglement distributions.  
Across a physically motivated parameter range, this classification accurately predicts the sign of many-body effects in the trimer.  
Our findings go beyond the Axilrod-Teller (AT) 
approximation, showing how AT predicts the wrong sign for the MB potential in the trimer. This may explain the failure of the AT potential to predict the attractive sign of the three-body effect in the para-hydrogen trimer at short range, compared to high level quantum chemistry methods~\cite{ibrahim2022three,hinde2008three,wind1996ab,ibrahim2022equation}.

Arbitrary arrangements of $N$ QDOs can be studied efficiently using our method, as extracting entanglement information only requires diagonalizing $3N \times 3N$ matrices, which scales as $\mathcal{O}(N^{3})$. 
As this scaling is identical to that required for computing bond energies, existing approaches can be readily augmented with the entanglement metrics introduced here.
Hence, while we have here focused on model systems, the same prescription, upon appropriate parameterization of the QDOs, can be extended to realistic materials. This would provide a direct link between their physical and chemical properties and the amount of entanglement shared by their constituents. Concepts from quantum information theory thus provide a novel path for shedding light on important chemical problems~\cite{gori2023second,lee2023evaluating,ding2020concept,ding2020correlation,molina2015quantum,boguslawski2013orbital,boguslawski2012entanglement,ding2022quantum}.

\section*{Acknowledgements}
CW acknowledges helpful discussion with Lewis W. Anderson.
Simulations were run
on the University of Oxford Advanced Research Computing (ARC) facility. JT is grateful for ongoing support through the Flatiron Institute, a division of the Simons Foundation. DJ acknowledges support by the European Union’s Horizon Programme (HORIZON-CL42021-DIGITALEMERGING-02-10) Grant Agreement 101080085 QCFD, the Cluster of Excellence ’Advanced Imaging of Matter’ of the Deutsche Forschungsgemeinschaft (DFG)- EXC 2056- project ID 390715994, and the Hamburg Quantum Computing Initiative (HQIC) project EFRE. The project is co-financed by ERDF of the European Union and by ‘Fonds of the Hamburg Ministry of Science, Research, Equalities and Districts (BWFGB)’. C.W. acknowledges support from the Engineering and Physical Sciences Research Council and IBM.
\newline

\section*{Declarations}
\begin{itemize}
\item There are no competing interests.
\item Code and data is freely available from the authors upon reasonable request.  
\end{itemize}

\bibliography{ref}

\clearpage
\newpage
\onecolumngrid

    \renewcommand{\theequation}{S\arabic{equation}}
    \renewcommand{\figurename}{Supplementary Figure}
    \setcounter{equation}{0}
    \setcounter{figure}{0}     

    \section*{Supplemental Material: Quantum Information Perspective on Many-Body Dispersive Forces}

\section{S1. The General QDO Hamiltonian applied in material simulation}
The general dipole order QDO Hamiltonian in units of $\hbar \omega_{0}$ is
\begin{equation}\label{FullQDOHamiltonianhbaromega}
\begin{gathered}
\hat{H}_{g}=\frac{1}{2}\sum^{N}_{\mu=1}|\boldsymbol{\hat{\mathcal{P}}}_{\mu}|^{2}+\frac{1}{2}\sum^{N}_{\mu=1} |\boldsymbol{\hat{\chi}}_{\mu}|^{2}\frac{\omega^{2}_{\mu}}{\omega^{2}_{0}}+\\
    \sum^{N}_{\mu >\xi}\frac{\omega_{\mu}\omega_{\xi}}{\omega^{2}_{0}}\sqrt{\alpha_{\mu}}\sqrt{\alpha_{\xi}} \boldsymbol{\hat{\chi}}_{\mu} e^{T}_{\mu}\mathcal{T}e_{\xi} \boldsymbol{\hat{\chi}}_{\xi},
\end{gathered}
\end{equation}
where  $\boldsymbol{\hat{\chi}}_{\mu}=\boldsymbol{\hat{r}}_{\mu}\sqrt{\omega_{0} m_{\mu}/\hbar}$, $\boldsymbol{\hat{\mathcal{P}}}_{\mu}=\boldsymbol{\hat{p}}_{\mu}/\sqrt{\hbar m_{\mu} \omega_{0}}$. Here $\omega_{0}$ is a reference frequency and $\alpha_{\mu}=q^{2}_{\mu}/(m_{\mu}\omega^{2}_{\mu}4\pi \epsilon_{0})$ is the polarizability of QDO $\mu$. 

The general dispersive binding energy is given by
\begin{equation}\label{FullQDObondbaromega}
E^{g}=\frac{3}{2}\sum^{N}_{\mu=1}\frac{\omega_{\mu}}{\omega_{0}}-\frac{1}{2}\sum^{3N}_{i=1}\sqrt{\lambda^{g}_{i}},
\end{equation}

where $\lambda^{g}_{i}$ are the eigenvalues of the matrix $V^{g}$ where $e^{T}_{\mu}V^{g}e_{\mu}=(\omega^{2}_{\mu}/\omega^{2}_{0})\mathds{1}_{3}$
and $e^{T}_{\mu}V^{g}e_{\xi}=\frac{\omega_{\mu}\omega_{\xi}}{\omega^{2}_{0}}\sqrt{\alpha_{\mu}}\sqrt{\alpha_{\xi}} \boldsymbol{\hat{\chi}}_{\mu}e^{T}_{\mu}\mathcal{T}e_{\xi}$, given $\mu \neq \xi$. Note that $e_{\mu}$ as well as $\mathcal{T}$ have been defined in the main-text. The general covariance matrix is thus computable in terms of the matrix $V^{g}$, as $\gamma^{g}_{0}=(V^{g})^{-1/2}\oplus (V^{g})^{1/2} $. Here $((V^{g})^{-1/2})_{ij}=\langle \hat{\chi}_{i}\hat{\chi}_{j} \rangle$ and 
$((V^{g})^{1/2})_{ij}=\langle \hat{\mathcal{P}}_{i}\hat{\mathcal{P}}_{j}\rangle$. The more general CM $\gamma^{g}_{0}$, replaces the CM in the main-text, when numerically studying arbitrarily parameterized QDOs. Extracting the necessary entanglement information follows the same procedure as defined in the main-text, but instead now using the elements of $\gamma^{g}_{0}$.

Building on the general formulation of the dipole-order QDO Hamiltonian in Eq.~\eqref{FullQDOHamiltonianhbaromega} and its associated bond energy expression, we now discuss how solving variants of this Hamiltonian have been effectively integrated into quantum chemical simulations of realistic molecular systems via the so-called MBD method. This method has been implemented in a range of electronic structure codes such as \textsc{Q-Chem}, \textsc{Quantum ESPRESSO}, and DFT-D4, as well as in standalone software packages~\cite{hermann2023libmbd}.

In the MBD method the dipole order QDO Hamiltonian is constructed on top of a given density functional and thus exhibits a high degree of versatility, depending on both the choice of density functional and the specific procedure used to determine the QDO parameters. In the MBD@SCS approach~\cite{ambrosetti2014long}, the QDO parameters \( \{\omega_{\mu}, m_{\mu}, q_{\mu}\} \) are computed to encode modified free-atom polarizability reference data, with atomic volumes scaled according to the Hirshfeld-partitioned electron density obtained from a DFT calculation~\cite{hermann2023libmbd}.

In this specifc MBD variant, the dipole interaction coupling matrix between QDOs \(\mu\) and \(\xi\), originally defined in Eq.~(1) of the main text, is modified to account for the finite spatial extent of the QDO charge densities:
\begin{equation}
    e^{T}_{\mu}\mathcal{T}^{\mathrm{scr}}e_{\xi} = \left(\mathrm{erf}(\zeta^{\mu,\xi}) - \Theta(\zeta^{\mu,\xi})\right) e^{T}_{\mu}\mathcal{T}e_{\xi} + 2\zeta^{2}\Theta(\zeta^{\mu,\xi})\frac{\boldsymbol{R}_{\mu \xi} \otimes \boldsymbol{R}_{\mu \xi}}{R^{5}_{\mu\xi}},
\end{equation}
where \(\Theta(\zeta^{\mu,\xi}) = \frac{2\zeta^{\mu,\xi}}{\sqrt{\pi}} e^{-(\zeta^{\mu,\xi})^{2}}\) and \(\zeta^{\mu,\xi} = R_{\mu\xi}/\sqrt{\sigma^{2}_{\mu} + \sigma^{2}_{\xi}}\). The parameters \(\sigma_{\mu}\) and \(\sigma_{\xi}\) represent the widths of the Gaussian charge densities associated with QDOs \(\mu\) and \(\xi\), respectively.

This modified interaction leads to the construction of a new screened dipole interaction matrix \(\mathcal{T}^{\mathrm{scr}}\) of dimension \(3N \times 3N\), which replaces \(\mathcal{T}\) in Eq.~\eqref{FullQDOHamiltonianhbaromega}. Since the quadratic form of the Hamiltonian is preserved, this modification acts only to appropriately regularize the dipole-dipole coupling at short range, while recovering the unscreened form in the long-range limit.

The MBD@SCS method has been benchmarked across a range of chemical datasets, notably improving the performance of the Perdew–Burke–Ernzerhof (PBE) functional~\cite{perdew1996generalized} for non-covalent interactions. In the S66 test set, the inclusion of MBD corrections reduces the mean absolute error in interaction energies to 0.3~kcal/mol, compared to 2.3~kcal/mol for the uncorrected PBE functional~\cite{ambrosetti2014long}.

Various versions of the MBD scheme, have demonstrated success across a wide range of simulated materials~\cite{distasio2014many,reilly2014role,reilly2013seamless,ambrosetti2014hard,reilly2015van}. For example, applying MBD to the aforementioned PBE functional yields improved predictions for the relative stabilities of oxalic acid polymorphs ~\cite{reilly2015van}, in agreement with experimental observations, unlike PBE alone. Moreover, the MBD approach has been shown to achieve accuracy comparable to diffusion quantum Monte Carlo  for computing the binding energies of \( C_{70} \) complexes \cite{hermann2017nanoscale}. 

Even for materials where standard MBD underperforms, such as with transition-metal dichalcogenides \cite{liu2020semiempirical,bryenton2023many} variants of the method, such as the MBD-NL \cite{hermann2020density}, have been developed to overcome the challenges posed by specific systems. This highlights the generality and minimal empiricism of the overall framework.


\section{S2. Axilrod-Teller formulae}

The AT correction, looked at in the main-text, can be further written for $N \geq 3$ \cite{donchev2006many} as a sum over triplets of QDOs,
\begin{equation}\label{atpot}
    \delta_{3}=\frac{-\alpha^{3}}{32}\mathrm{tr}(\mathcal{T}^{3})=
    \frac{-9 \alpha^{3}}{16}\sum^{N}_{\o >\xi > \mu=1} \frac{[1+3\mathrm{cos}(\theta_{\mu \xi})\mathrm{cos}(\theta_{\mu \o})\mathrm{cos}(\theta_{\xi \o})]}{R^{3}_{\mu \xi}R^{3}_{\mu \o}R^{3}_{\xi \o}}.
\end{equation}
The angles $\theta_{\mu \xi}$ is given by,
\begin{equation}
\mathrm{cos}(\theta_{\mu \xi})=\frac{-R^{2}_{\mu \xi}+R^{2}_{\mu \o}+R^{2}_{\xi \o}}{2R_{\mu \o}R_{\xi \o}}.
\end{equation}

\section{S3. Properties of the ground state CM}\label{CMmathds}

\textit{Lemma. 1}\label{lem1} $(V^{1/2})_{ii} \leq 1 \: \forall i \in \{1,2,...,3N\}$.

\textit{Proof.} 
The matrix $V$ is a symmetric matrix, $V=V^{T}$, with positive eigenvalues $\lambda$. It is therefore diagonalized by a orthogonal matrix, $O$ ($OO^{T}=\mathds{1}$). The diagonal elements of the matrix square root of $V$ are given by
\begin{equation}
(V^{1/2})_{ii}=\sum^{3N}_{j=1}O^{2}_{ij}\sqrt{\lambda_{j}}, \quad{} \sum^{3N}_{j=1}O^{2}_{ij}=1.
\end{equation}
The diagonal elements of $V^{1/2}$ are given by a \textit{convex combination} of the square root of the  eigenvalues of $V$. We apply Jensen's inequality \cite{hansen2003jensen}, which states that 
\begin{equation}
\left(\sum^{3N}_{j=1}O^{2}_{ij}\sqrt{\lambda_{j}}\right)^{2} \leq \sum^{3N}_{j=1}O^{2}_{ij}\left(\sqrt{\lambda_{j}}\right)^{2},
\end{equation}
which gives the result, as by definition,
\begin{equation}
\sum^{3N}_{j=1}O^{2}_{ij}\lambda_{j}=V_{ii}=1.
\end{equation}

\textit{Lemma. 2} The dispersive binding energy  is always \textit{net} attractive \label{lem2}, i.e $E \geq 0$.

\textit{Proof.} This follows from lemma. 1,  as the bond energy is
\begin{equation}
    E=\frac{3}{2}N-\frac{1}{2}\sum^{3N}_{i=1}(V^{1/2})_{ii},
\end{equation}
where $E \geq 0$, directly follows from the fact that $(V^{1/2})_{ii} \leq 1 \forall i \in \{1,..,3N\}$.

\textit{Lemma. 3}\label{lem3} $(V^{-1/2})_{ii} \geq [(V^{1/2})_{ii}]^{-1}, \: \forall i \in \{1,..,3N\}.$\label{lem3}

\textit{Proof.} 
As $V^{1/2}$ is the unique symmetric positive definite matrix such that $(V^{1/2})^{2}=V$ \cite{horn2012matrix}, there is a unique symmetric positive matrix $C$ where $V^{1/2}=CC$ and thus also $V^{-1/2}=(C^{-1})^{2}$. 
Note that as $C$ is a symmetric matrix, $C_{ij}=C_{ji}$ and $(C^{-1})_{ij}=(C^{-1})_{ji}$. We can thus employ the  Cauchy-Schwartz inequality to complete the proof,
\begin{equation}
 \underbrace{\left(\sum^{3N}_{j=1}C_{ij}(C^{-1})_{ji}\right)^{2}}_{=1} \leq \underbrace{\left(\sum^{3N}_{j=1}C_{ij}^{2}\right)}_{(V^{1/2})_{ii}}\underbrace{\left(\sum^{3N}_{j=1}(C^{-1})^{2}_{ji}\right)}_{{(V^{-1/2})_{ii}}}.
\end{equation}

\section{S4. Symplectic diagonalization}

\begin{figure}
    \centering    \includegraphics{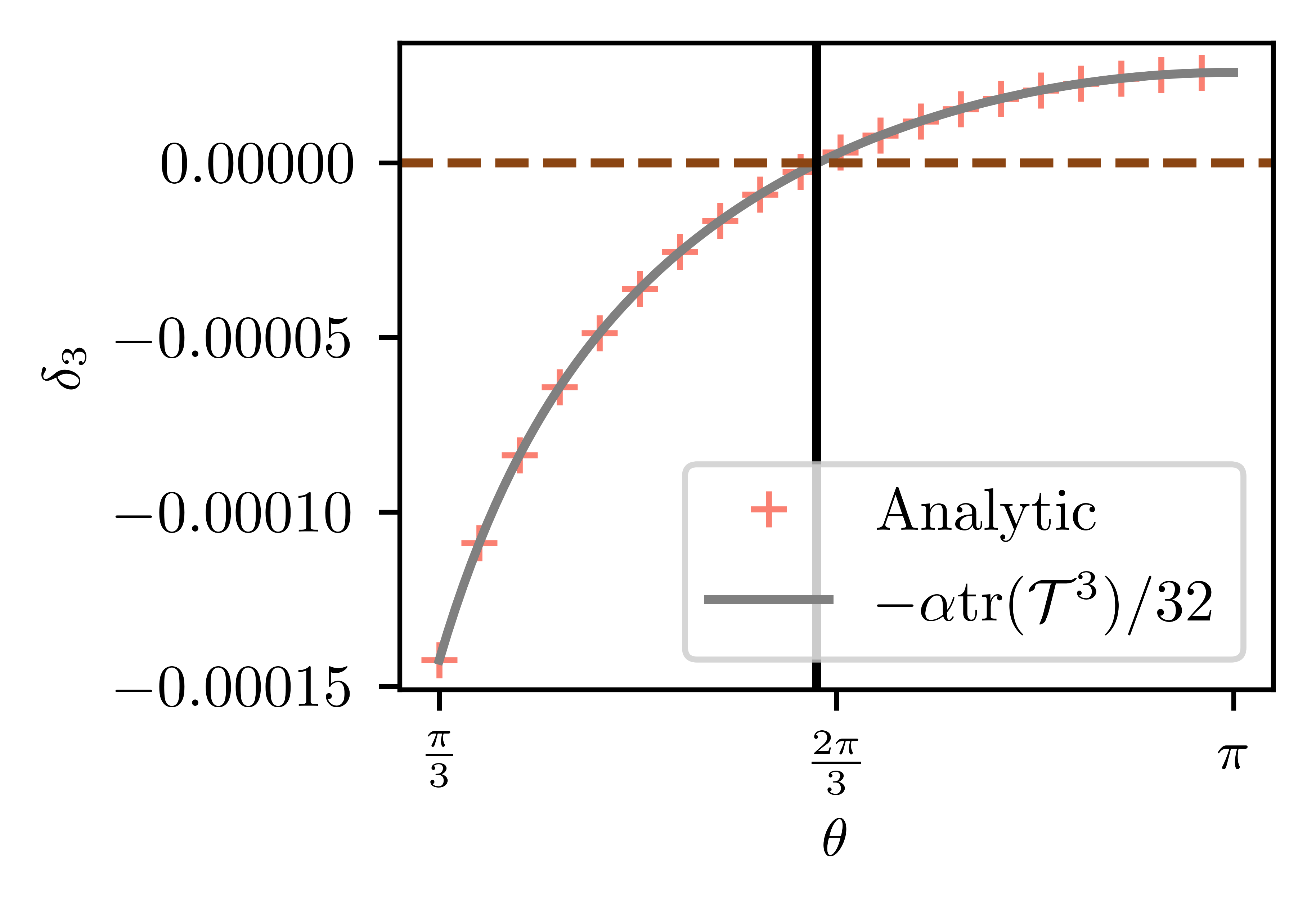}
    \caption{AT potential in the trimer at $\rho=2.60$ (see main-text for description). The orange crosses show $\delta_{3}$ computed via the R.H.S of Eq. \eqref{atpot}, whereas the solid grey curve shows $\delta_{3}$ computed by the L.H.S of Eq. \eqref{atpot}}
    \label{ATequiv}
\end{figure}
A general quadratic bosonic Hamiltonian can always be written as
\begin{equation}
\hat{H}=\frac{1}{2}\boldsymbol{\hat{q}}^{T}H\boldsymbol{\hat{q}}, \quad{} H >0,
\end{equation}
where $\boldsymbol{\hat{q}}=(\hat{\chi}_{1},...,\hat{\chi}_{d},\hat{\mathcal{P}}_{1},...,\hat{\mathcal{P}}_{d})^{T}$ and $d$ is the number of modes in the system. 
The ground state energy can be found variationally in terms of the CM as
\begin{equation}\label{gsenergy}
E_{0}=\mathrm{inf}_{\gamma} \frac{1}{4}\rm{tr}(\gamma H), \quad{} \mathrm{with} \quad{} \gamma + i\sigma  \geq 0,
\end{equation}
where the matrix $\sigma$ encodes the commutation relations, 
\begin{equation}
\sigma=\begin{bmatrix} 0 & \mathds{1} \\-\mathds{1} & 0  \end{bmatrix}.
\end{equation}
The constraint $\gamma + \rm{i} \sigma
\geq 0$, enforces the Heisenberg uncertainty principle.
A matrix which preserves $\sigma$ and therefore the commutation relations under a transformation is a symplectic matrix \cite{serafini2017quantum}, belonging to the symplectic matrix group,
\begin{equation}
    S \sigma S^{T}=\sigma \quad{} S\in \mathcal{S}_{p}(\mathrm{dim}(2d,\mathds{R}).
\end{equation}
Williamson's theorem \cite{williamson1936algebraic} states there is always a symplectic matrix which brings $H$ into diagonal form,
\begin{equation}\label{symps}
    SHS^{T}= \begin{pmatrix}
\epsilon_{1} & 0 & \cdots & 0 & 0 \\
0 & \epsilon_{1} & \cdots & 0 & 0 \\
\vdots & \vdots & \ddots & \vdots & \vdots \\
0 & 0 & \cdots & \epsilon_{d} & 0 \\
0 & 0 & \cdots & 0 & \epsilon_{d}
\end{pmatrix},
\end{equation}
where $\epsilon_{i}$ are the symplectic eigenvalues of $H$, each repeated once along the diagonal. The ground state CM can be written as $\gamma_{0}=S^{T}S$ \cite{schuch2006quantum}.
From the unitary representation of the symplectic matrix, which brings $H$ into diagonal form, the ground state energy is given by the sum of the unique symplectic eigenvalues \cite{schuch2006quantum,serafini2017quantum},
\begin{equation}\label{symplecticenergy}
    E_{0}=\frac{1}{2}\sum^{d}_{i=1}\epsilon_{i}.
\end{equation}
\section{S5. The Tangle in a pure two mode state}\label{tmt}
The two mode ($d=2$) quadratic Hamiltonian, coupled in position operators is given by \begin{equation}\label{twomodeham}
    \hat{H}=\frac{1}{2}\left(\hat{\chi}^{2}_{1}+\hat{\chi}^{2}_{2}+\hat{\mathcal{P}}^{2}_{1}+\hat{\mathcal{P}}^{2}_{2}\right)+\kappa\hat{\chi}_{1}\hat{\chi}_{2}, \quad{} |\kappa| <1.
\end{equation}
In order for the positive definite constraint on the Hamiltonian matrix to be obeyed, $|\kappa|<1$.
The Hamiltonian in Eq. \eqref{twomodeham} is equivalent to the two mode Hamiltonian $\hat{H}_{ij}$, in the main-text, for $\kappa=\alpha \mathcal{T}_{ij}$.

The symplectic spectrum $\{\epsilon_{+},\epsilon_{-},\epsilon_{+},\epsilon_{-} \}$  is given by,
\begin{equation}\label{explicitysymp}
    \epsilon_{\pm}=\sqrt{1\pm \kappa}.
\end{equation}
The two mode ground state correlation matrix, is given by
\begin{equation}\label{explicitytwomodecorrel}
\gamma_{0}=
\frac{1}{2}\begin{pmatrix}
\epsilon^{-1}_{+}+\epsilon^{-1}_{-}&\epsilon^{-1}_{+}-\epsilon^{-1}_{-}\\\epsilon^{-1}_{+}-\epsilon^{-1}_{-}&\epsilon^{-1}_{+}+\epsilon^{-1}_{-}
\end{pmatrix} 
\oplus 
\begin{pmatrix} 
\epsilon_{+}+\epsilon_{-}&\epsilon_{+}-\epsilon_{-}\\\epsilon_{+}-\epsilon_{-}&\epsilon_{+}+\epsilon_{-}
\end{pmatrix},
\end{equation}
where $2\langle\hat{\chi}^{2}_{1}\rangle=2\langle\hat{\chi}^{2}_{2}\rangle=\epsilon^{-1}_{+}+\epsilon^{-1}_{-}$, $2\langle\hat{\chi}_{1}\hat{\chi}_{2}\rangle=\epsilon^{-1}_{+}-\epsilon^{-1}_{-}$, $2\langle\hat{\mathcal{P}}^{2}_{1}\rangle=2\langle\hat{\mathcal{P}}^{2}_{2}\rangle=\epsilon_{+}+\epsilon_{-}$ and 
$2\langle\hat{\mathcal{P}}_{1}\hat{\mathcal{P}}_{2}\rangle=\epsilon_{+}-\epsilon_{-}$.

From the definition of the Gaussian tangle in the main-text and using the form of the two mode CM defined in Eq. \eqref{explicitytwomodecorrel},
\begin{equation}\label{simplifiedexp}
    \tau_{G}(1)=\tau_{G}(1:2)=\frac{1}{4}\left( \sqrt{\frac{\epsilon_{+}}{\epsilon_{-}}}-1\right)^{2},
\end{equation}
from which the definition of $\tau_{G}^{\mathrm{ref}}(i:j)$ follows.

\section{S6. Mixed state Pairwise Tangle}\label{tmt}
The pairwise tangle, $\tau_{G}(i:j)$, quantifying the entanglement shared between modes $i$ and $j$ in a mixed two mode state, is elaborated on here. We define the following quantities, which uniquely capture the two mode entanglement; $a^{ii},b^{jj},c^{ij}_{+},c_{-}^{ij}$, with $c^{ij}_{+} \geq |c_{-}^{ij}|$ . These are required to satisfy the following equalities: $a^{ii}=\sqrt{\langle \hat{\chi}^{2}_{i}\rangle\langle \hat{P}_{i}^{2} \rangle}$, $b^{jj}=\sqrt{\langle \hat{\chi}^{2}_{j}\rangle\langle \hat{P}_{j}^{2} \rangle}$, $c^{ij}_{+}c_{-}^{ij}=\langle \hat{\chi}_{i}\hat{\chi}_{j}\rangle\langle \hat{P}_{i}\hat{P}_{j} \rangle$ and $(a^{ii}b^{jj}-(c^{ij}_{+})^{2})(a^{ii}b^{jj}-(c^{ij}_{-})^{2})=((\langle \hat{P}_{i}\hat{P}_{j} \rangle)^{2}-\langle \hat{P}_{i}^{2} \rangle\langle \hat{P}_{j}^{2} \rangle)((\langle \hat{\chi}_{i}\hat{\chi}_{j}\rangle)^{2}-\langle\hat{\chi}^{2}_{i}\rangle\langle\hat{\chi}^{2}_{j}\rangle)$. In addition the following inequalities must be satisfied: $a^{ii} \geq 1$, $b^{jj} \geq 1$ (which follows from lemma. 3), $a^{ii}b^{jj}-(c^{ij}_{\pm})^{2} \geq 1$ and $(a^{ii}b^{jj}-(c^{ij}_{+})^{2})(a^{ii}b^{jj}-(c^{ij}_{-})^{2})+1 \geq (a^{ii})^{2}+(b^{jj})^{2}+2c^{ij}_{+}c^{ij}_{-}$. 
For notational convenience we will write $a^{ii},b^{jj},c^{ij}_{+},c_{-}^{ij}$ as $a,b,c_{+},c_{-}$, for here onwards. The pairwise tangle is then given by \cite{hiroshima2007monogamy}, 
\begin{equation}
    \tau_{G}(i:j)=f(\mathrm{min}_{0 \leq \phi < 2\pi} m(\phi)),
\end{equation}
where recall the definition of $f(x)$ given in the main-text. Here
\begin{equation}
    m(\phi)=h^{2}_{1}(\phi)/h_{2}(\phi),
\end{equation}
\begin{equation}
    h_{1}(\phi)=\xi_{-}+\sqrt{\eta^{*}}\cos(\phi),
\end{equation}
\begin{equation}
     h_{2}(\phi)=2(ab-c^{2}_{-})(a^{2}+b^{2}+2c_{+}c_{-})-\\(\zeta/\sqrt{\eta^{*}})\cos(\phi)+
(a^{2}-b^{2})\sqrt{1-\xi^{2}_{+}/\eta^{*}}\sin(\phi)
\end{equation}
\begin{equation}
    \xi_{\pm}=c_{+}(ab-c^{2}_{-})\pm c_{-}
\end{equation}
\begin{equation}
    \eta^{*}=[a-b(ab-c^{2}_{-}][b-a(ab-c^{2}_{-}])]
\end{equation}
and 
\begin{equation}
    \zeta=2abc^{3}_{-}+(a^{2}+b^{2})c_{+}c^{2}_{-}+[a^{2}+b^{2}-2a^{2}b^{2}]c_{-}\\ -ab(a^{2}+b^{2}-2)c_{+}.
\end{equation}
\section{S7. Three mode Hamiltonian }\label{tmt3}
\begin{figure}
    \centering
    \includegraphics{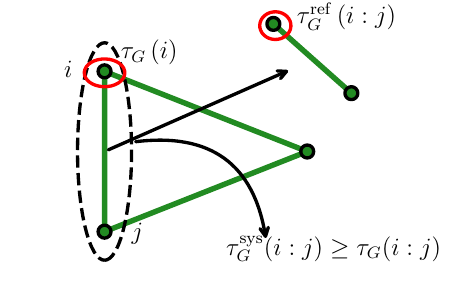}
    \caption{Coupling graph of a $3$ mode Hamiltonian. The red circle illustrates the bipartition between mode $i$ and the other two modes, where $\tau_{G}(i)$ measures the entanglement across this bipartition and limits the strength of the pairwise tangle between modes $i$ and $j$, to be no greater then $\tau^{\mathrm{sys}}_{G}(i:j)$. The reference tangle, given by $\tau^{\mathrm{ref}}_{G}(i:j)$, is the Gaussian tangle between $i$ and $j$ in the  ground state of the two mode Hamiltonian, $\hat{H}_{ij}$.}
    \label{diagthreemode}
\end{figure}
Here we will  extend the Hamiltonian looked at in Eq. \eqref{twomodeham}. The three mode Hamiltonian we will consider here is
\begin{equation}\label{threemode}
\hat{H}=\frac{1}{2}\left(\hat{\chi}^{2}_{1}+\hat{\chi}^{2}_{2}+\hat{\chi}^{2}_{3}+\hat{\mathcal{P}}^{2}_{1}+\hat{\mathcal{P}}^{2}_{2}+\hat{\mathcal{P}}^{2}_{3}\right)+\kappa\hat{\chi}_{1}\hat{\chi}_{2}+\kappa\beta \hat{\chi}_{1}\hat{\chi}_{3}+\kappa \hat{\chi}_{2}\hat{\chi}_{3},  \quad{} |\kappa|,\beta \in [0,1].
\end{equation}
We investigate how the interaction with the third mode, modifies the shareability of the entanglement between modes 1 and 2. In the tripartite Hamiltonian modes 1 and 2 are coupled by the parameter $\kappa$. Modes 2 and 3 have an identical coupling strength. In comparison the coupling between modes 1 and 3 varies with the parameter $\beta$. When $\beta=0$, modes 1 and 3 share no coupling and when $\beta=1$, each mode in the system interacts identically with the other two modes.
\begin{figure}
    \centering
\includegraphics{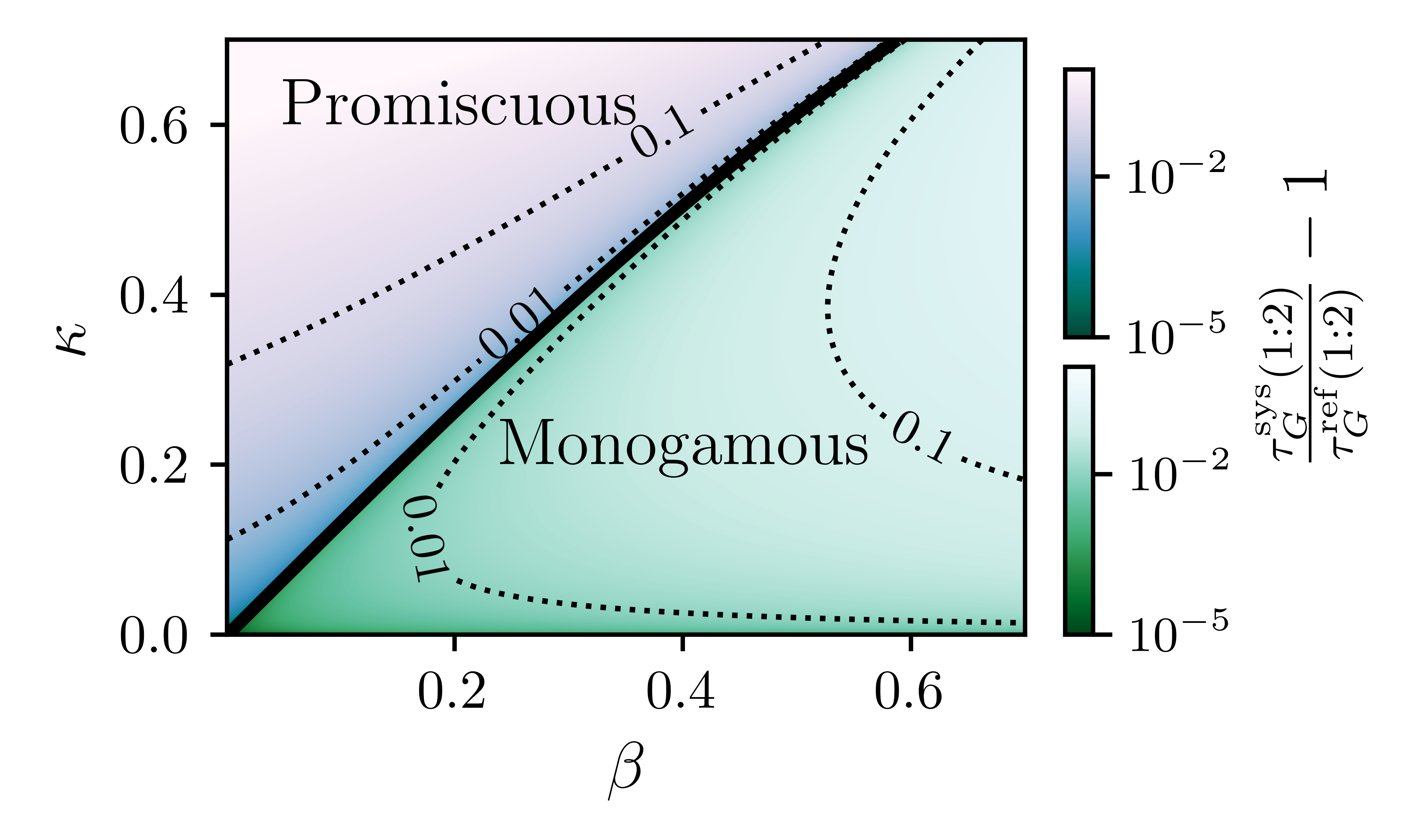}
    \caption{Behavior of  $\tau^{\mathrm{sys}}_{G}(1:2)/\tau^{\mathrm{ref}}_{G}(1:2)-1$ as a function of $\kappa$ and $\beta$, see Eq. \eqref{threemode} for definitions of these parameters. The monogamy/promiscuity boundary is shown by the solid black line, below which the entanglement behaves monogamously and above which, promiscuously. Both regions are shown as heat-maps, plotted on a log-scale.   }
    \label{threemodeheatmap}
\end{figure}

The reference tangle between modes 1 and 2, is given by $\tau^{\mathrm{ref}}_{G}(1:2)$ and is computable in terms of $\kappa$ via 
the definition of the tangle in the two mode state, derived in Eq. \eqref{simplifiedexp}. See also Fig. \ref{diagthreemode} for an illustrated definition, in a three mode Hamiltonian.
In comparison in the tripartite Hamiltonian, the maximal pairwise tangle is given by $\tau^{\star}_{G}(1:2)$. Note that due to the symmetry of the tripartite system, the entanglement and correlations shared between modes 1 and 2 must be identical to that shared between modes 2 and 3.
Fig. \ref{threemodeheatmap} shows a heatmap of $\tau^{\mathrm{sys}}_{G}(1:2)/\tau^{\mathrm{ref}}_{G}(1:2)-1$, as a function of positive $\kappa$ and $\beta$. This is shown for 2000 $\beta$ values, where $\beta \in \{0,0.7\}$ and 2000 $\kappa$ values, for
 $\kappa \in \{0.001,0.7\}$. In the limit where $\kappa \rightarrow 0$, $\tau^{\mathrm{sys}}_{G}(1:2)/\tau^{\mathrm{ref}}_{G}(1:2)\rightarrow 1$.

For $\beta>0$, at weak couplings the pairwise entanglement is restricted. As a function of increasing $\kappa$, the pairwise entanglement shifts toward behaving promiscuously. The solid black line shows where $\tau^{\mathrm{sys}}_{G}(1:2)=\tau^{\mathrm{ref}}_{G}(1:2)$, splitting the parameter space into its respective monogamous and promiscuous regions. For strong coupling i.e at large $\kappa$ and for most of the shown $\beta$ values, the pairwise entanglement behaves promiscuously. To better understand the change from monogamous to promiscuous behavior as a function of coupling strength, we look at a reduced Hilbert space description of the three mode Hamiltonian.

\begin{figure}
    \centering
\includegraphics{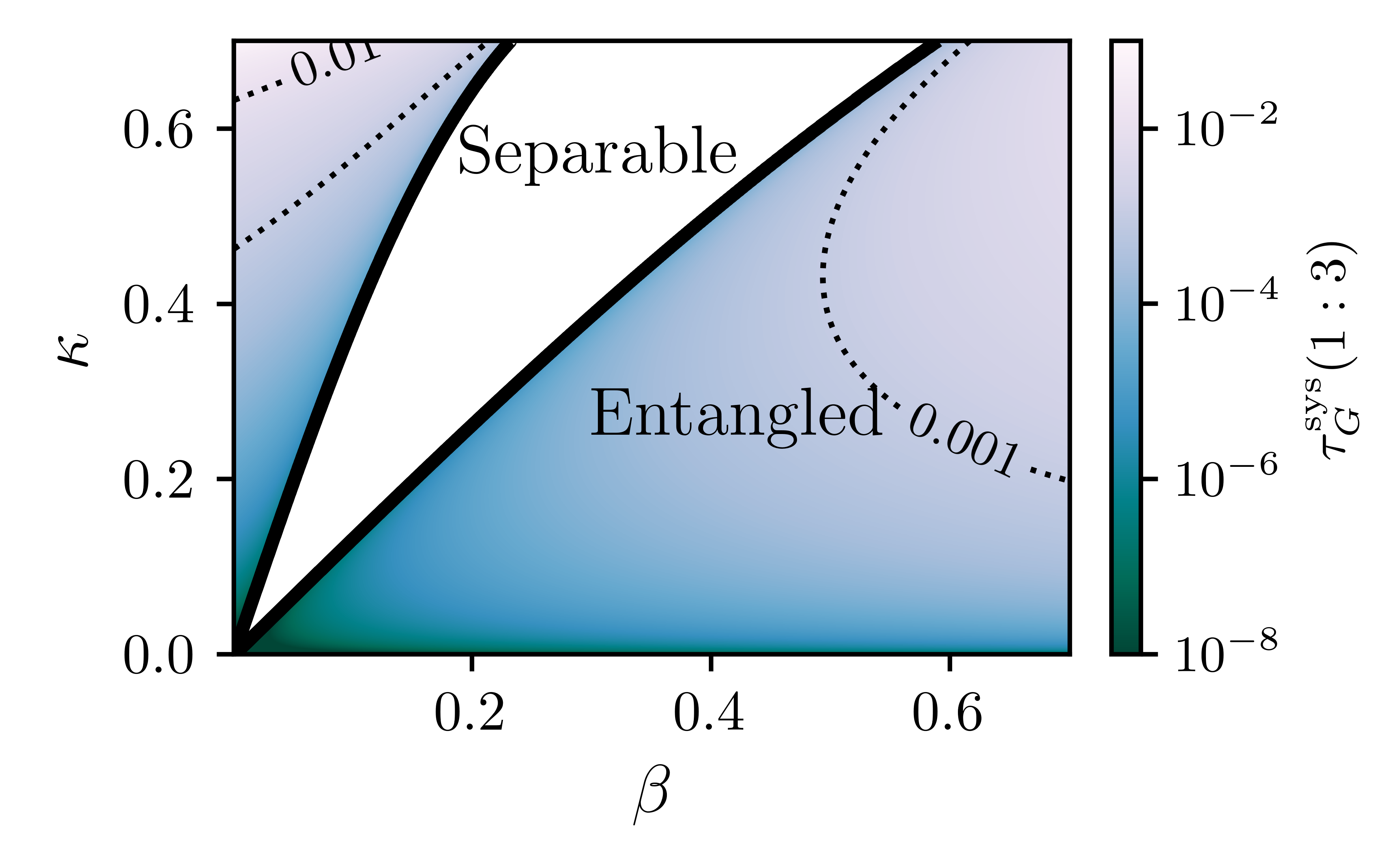}
    \caption{ Heatmap of $\tau^{\mathrm{sys}}_{G}(1:3)$, for the same setup and thus same $\beta$ and $\kappa$ values as looked in Fig. \ref{threemodeheatmap}. The are $\kappa$ and $\beta$ values, shown by the colorless region, show where $\langle \hat{\chi}_{1}\hat{\chi}_{3}\rangle\langle \hat{P}_{1}\hat{P}_{3} \rangle>0$, with the solid black lines showing where $\langle \hat{\chi}_{1}\hat{\chi}_{3}\rangle\langle \hat{P}_{1}\hat{P}_{3} \rangle=0$. From the PPT criterion given in the main-text, modes $1$ and $3$ are guaranteed to share no entanglement for this region of the phase space. }
    \label{tripartiteheatmap2}
\end{figure}

\section{S8. Low dimensional approximation}
Gaussian states make up a subsection of continuous variable quantum states and thus have an underlying Hilbert space description. A quantum system is called a continuous-variable system
when it has an infinite-dimensional Hilbert space. A general $d$ mode system can be described by a Hilbert space with a tensor product structure, $\mathcal{H}^{\otimes d}=\otimes^{d}_{i=1}\mathcal{H}_{i}$.

In addition to the quadrature operators, we can also work with bosonic field operators, which are written in terms of the quadrature field operators as
\begin{equation}
  \hat{\chi}_{i}  =  \frac{(\hat{a}^{\dagger}_{i}+\hat{a}_{i})}{\sqrt{2}} \quad{}\hat{\mathcal{P}}_{i} = \frac{\mathrm{i}(\hat{a}^{\dagger}_{i}-\hat{a}_{i})}{\sqrt{2}}.
\end{equation}
A multimodal Hilbert space  $\mathcal{H}$ is spanned by the basis $\{\ket{n}
\}^{\infty}_{n=0}$, known as the Fock or number basis, given by the eigenstates of the number operator $\hat{n} = \hat{a}^{\dagger}\hat{a}$, where $\hat{n}\ket{n}=n\ket{n}$. 
Over the infinite dimensional Fock space, the actions of the lowering and raising Bosonic field operators are as follows; $\hat{a}\ket{0}=0$ and $\hat{a}\ket{n}=\sqrt{n}\ket{n-1}, \forall n\geq 1$ and $\hat{a}^{\dagger}\ket{n}=\sqrt{n+1}\ket{n+1}$. 

Transforming between the quadrature vector  $\boldsymbol{\hat{q}}=(\hat{\chi}_{1},...,\hat{\chi}_{d},\hat{\mathcal{P}}_{1},...,\hat{\mathcal{P}}_{d})^{T}$ and the vector of ladder operators $ \hat{\boldsymbol{a}}=(\hat{a}_{1},\hat{a}_{2},..,\hat{a}_{d},\hat{a}^{\dagger}_{1},\hat{a}^{\dagger}_{2},..,\hat{a}^{\dagger}_{d})^{T}$ is given via a unitary transform, $L_{(c)}$. Thus the  ladder operator representation of the ground state CM is given as follows \cite{adesso2014continuous},
\begin{equation}\label{complextransform}
\gamma_{0} \rightarrow \gamma^{(c)}_{0}=L_{(c)}\gamma_{0}L^{\dagger}_{(c)}, \quad{} L_{(c)}=\frac{1}{\sqrt{2}}\begin{pmatrix}
    \mathds{1}& \rm{i} \mathds{1}\\
    \mathds{1}& -\rm{i} \mathds{1}
\end{pmatrix}.
\end{equation}

The energy gap of the system is given by $\mathrm{min}(\epsilon_{i})$, where $\epsilon_{i}$
$\forall i \in \{1,.,d\}$ are the symplectic eigenvalues in Eq. \eqref{symps} ~\cite{schuch2006quantum}. When $\mathrm{min}(\epsilon_{i})=0$ the system is said to be critical~\cite{schuch2006quantum}. Considering the two mode example with Hamiltonian in Eq. \eqref{twomodeham}, the critical point is approached as $\kappa \rightarrow 1$, as here $\epsilon_{-} \rightarrow 0$. 

Transforming the analytic expression for the two-mode ground-state CM in Eq.~\eqref{explicitytwomodecorrel} via the unitary transformation in Eq.~\eqref{complextransform} yields diagonal CM elements:
\[
\mathrm{diag}(\gamma^{(c)}_{0}) = 0.5\left(\epsilon_{+} + \epsilon_{+}^{-1} + \epsilon_{-} + \epsilon_{-}^{-1}\right)\mathds{1}_{4}.
\]
Hence, as $\kappa \rightarrow 1$, $(\gamma^{(c)}_{0})_{ii} = (\gamma^{(c)}_{0})_{ii} \rightarrow \infty$, where $i \in \{1,2,3,4\}$. This shows that in the critical limit, the ground state expectation value of the number operator diverges in the two-mode state, while remaining finite away from criticality. Ref.~\cite{anderson2022coarse} investigated the approximation error in the correlation energy for the Hamiltonian in Eq.~\eqref{twomodeham} arising from a finite truncation of the Hilbert space and found the error to be negligible, away from the critical point.

Using the truncated Fock state approach (see \cite{anderson2022coarse} for details), a quadratic $
d$ mode Hamiltonian with coupling only between position coordinates — using only the first two Fock states i.e $\ket{0}$ and $\ket{1}$ — leads to the following  $d$ \textit{qubit} Hamiltonian,
\begin{equation}\label{qubit}
    \hat{H}^{d}_{\mathrm{qub}}=\frac{1}{2}\sum^{d}_{i=1}\left(2\mathds{1}^{i}_{2}-\sigma^{i}_{z}\right)+\sum^{d}_{j> i}w_{ij}\sigma^{i}_{x}\otimes\sigma^{j}_{x}.
\end{equation}
Here $\sigma^{i}_{z}$ ($\sigma^{i}_{x}$) is the $z$ ($x$) Pauli matrix, acting on the $i$th qubit, $w_{ij}$ is determined by the coupling strength between modes and $\otimes$ denotes the tensor product. By introducing the qubit Hamiltonian in Eq. \eqref{qubit}, we will be able to contrast the monogamy/promiscuity behavior in the infinite dimensional Gaussian ground state, with the monogamy/promiscuity behavior in a two state approximation of the infinite Hilbert space.
\subsection{A. The qubit tangle}
Consider a $d$ qubit system. The  $d$ qubit ground state density matrix is a $2^{d} \times 2^{d}$ matrix, denoted by $\hat{\rho}^{d}$. The superscript on the qubit density matrix, distinguishes it as a finite dimensional matrix, given that $d \not=\infty$. This is in comparison to the infinite dimensional density matrices, indirectly considered in the main-text. 

The tangle between qubit $i$ and $d-1$ qubits, i.e for a bipartition of the form $i:\underbrace{k,l,j,..,m}_{d-1}$, is given in terms of the concurrence, another entanglement monotone \cite{coffman2000distributed,osborne2006general} as
\begin{equation}\label{qubitang}
    \tau(i)=\mathcal{C}^{2}(i)=4\mathrm{Det}(\hat{\rho}^{d}_{i}),
\end{equation}
where $\hat{\rho}^{d}_{i}$ is the reduced density matrix of the $i$th qubit, computed via tracing out modes $k,l,j,..,m$. The monogamy inequality is then given as
\begin{equation}
      \tau(i) \geq \sum^{d}_{j=1}\tau(i:j), \forall j \not = i,
\end{equation}
where $\tau(i:j)$ denotes the pairwise tangle between qubits $i$ and $j$. 

Let $\hat{\rho}^{d}_{ij}$ denote the two qubit reduced density matrix of qubits $i$ and $j$, tracing over modes $k,l,..,m$, in the $d$ qubit system. Defining,
\begin{equation}
    (\hat{\rho}_{ij}^{d})^{'}= (\sigma_{y} \otimes \sigma_{y})(\hat{\rho}_{ij}^{d})^{\star}(\sigma_{y} \otimes \sigma_{y}),
\end{equation}
where $\sigma_{y}$ is the $y$ Pauli gate, $\sigma_{y}=\begin{pmatrix}
    0 & -\mathrm{i} \\ \mathrm{i} & 0 
\end{pmatrix}$ and $(\hat{\rho}_{ij}^{d})^{\star}$ is the complex conjugate of $\hat{\rho}_{ij}^{d}$. By computing 
\begin{equation}
\lambda^{'}_{i}=\sqrt{\mathrm{eig}_{i}\left((\hat{\rho}_{ij}^{d})^{\star}(\hat{\rho}_{ij}^{d}){'}\right)}, \quad{}  \forall i \in \{1,2,3,4\},
\end{equation}
where $\lambda^{'}_{1} \geq \lambda^{'}_{2} \geq \lambda^{'}_{3} \geq \lambda^{'}_{4}$. The pairwise tangle between qubits $i$ and $j$ is given as \cite{coffman2000distributed},
\begin{equation}\label{pairwisequbittangle}
    \tau(i:j)=\left[\mathrm{max}\{\lambda^{'}_{1}-\lambda^{'}_{2}-\lambda^{'}_{3}-\lambda^{'}_{4},0\}\right]^{2}.
\end{equation}

\subsection{B. Low dimensional approximation of the three mode Hamiltonian}

\begin{figure}
    \centering
    \includegraphics{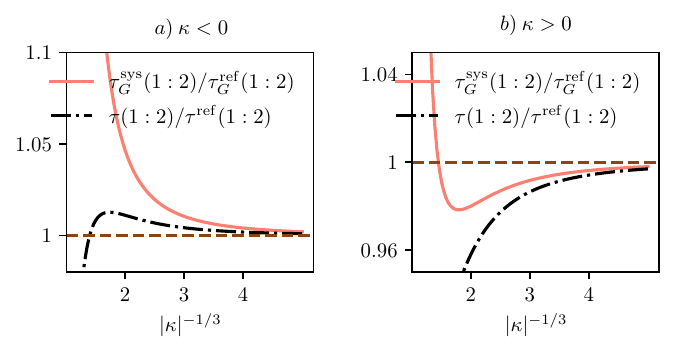}
    \caption{ The solid orange curve is the tangle computed from solving Eq. \eqref{threemode} exactly. The dashed dotted black line, follows from the truncated Fock state approximation, with Hamiltonian given in Eq. \eqref{qubit} and $d=3$. Below the horizontal dashed line shows where the pairwise entanglement between 1 and 2 behaves monogamously whereas above shows the promiscuous behaviour. }
    \label{tanglequbit}
\end{figure}


Here we solve for the qubit approximation of Eq. \eqref{threemode}, $\hat{H}^{d=3}_{\mathrm{qub}}$, given in Eq. \eqref{qubit}. When  $d=2$, Eq. \eqref{qubit} is a two qubit Hamiltonian, with $w_{12}=\kappa$. The general ground state density matrix is a $2^{d} \times 2^{d}$ matrix, denoted by $\hat{\rho}^{d}$. The  $4\times 4$ ground state density matrix of $\hat{H}^{d=2}_{\mathrm{qub}}$  is written as $\hat{\rho}^{2}$. The tangle between qubits 1 and 2, can be straightforwardly computed from the reduced density matrix, tracing out qubit 2, giving the $2 \times 2$ reduced density matrix $\hat{\rho}^{2}_{2}$. From Eq. \eqref{qubitang}  the tangle between qubits  1 and 2 is given by,
\begin{equation}
\tau^{\mathrm{ref}}(1:2)=4\mathrm{Det}(\hat{\rho}^{2}_{1})=4\mathrm{Det}(\hat{\rho}^{2}_{2})
\end{equation}
This is analogously labeled to the Gaussian tangle in the two mode pure state, denoted here as  $\tau_{G}^{\mathrm{ref}}(1:2)$. 
Given $d=3$, $\hat{H}^{3}_{\mathrm{qubit}}$ is a three qubit Hamiltonian, where $w_{12}=w_{23}=\kappa$ and $w_{13}=\beta\kappa$. The ground state density matrix is given by $\hat{\rho}^{3}$. The monogamy inequality for distributed qubit entanglement, is given in terms of the tangle as $\tau(1) \geq \tau(1:2)+\tau(1:3)$, where $\tau(1)=4\mathrm{Det}(\hat{\rho}^{3}_{1})$ and $\hat{\rho}^{3}_{1}$ is a $2 \times 2$ matrix, found by tracing out qubits 2 and 3. The pairwise tangle $\tau(1:2)$ and  $\tau(1:3)$ in the three qubit state is computed from Eq. \eqref{pairwisequbittangle}.

In Fig. \ref{tanglequbit}, we show how $\tau^{\mathrm{ref}}_{G}(1:2)$ and $\tau^{\mathrm{ref}}(1:2)$ are modified, due to the tripartite interaction,  setting $\beta=1/4$, and for negative and positive coupling $\kappa$, i.e $a)$ and $b)$. Note that the $x-$axis in Fig. \ref{tanglequbit} is written in terms of $|\kappa|^{-1/3}$, which is done in order for a convenient comparison with later chapters. From  Fig. \ref{tanglequbit} $a)$, both $\tau^{\mathrm{sys}}_{G}(1:2)>\tau^{\mathrm{ref}}_{G}(1:2)$ and $\tau(1:2)>\tau^{\mathrm{ref}}(1:2)$, for $|\kappa|^{-1/3} \leq  2$. 
In comparison for $0 
\leq \kappa^{-1/3} \leq 2$, shown in $b)$, both $\tau^{\mathrm{sys}}_{G}(1:2)<\tau^{\mathrm{ref}}_{G}(1:2)$ and $\tau(1:2)<\tau^{\mathrm{ref}}(1:2)$. Thus for a three mode (qubit) state, with positive intermodal couplings, mode (qubit) 3 restricts the shared tangle between modes (qubits) 1 and 2. In $b)$, for increased coupling strength, a sign change occurs in  $\tau^{\mathrm{sys}}_{G}(1:2)-\tau^{\mathrm{ref}}_{G}(1:2)$ but not in $\tau(1:2)-\tau^{\mathrm{ref}}(1:2)$. The pairwise tangle between qubits becomes increasing restricted as a function of coupling strength.


In the limit of $\kappa \rightarrow 0$ for negative inter-modal (inter-qubit) coupling, the pairwise entanglement behaves promiscuously. Conversely in the 
of $\kappa \rightarrow 0$ for positive inter-modal (inter-qubit) coupling, the pairwise entanglement behaves monogamously, where here the pairwise tangle in both the multi-qubit and multimode ground state is suppressed.
We thus find that in the non-interacting limit, truncating the bosonic Hilbert space to  a two-level system, captures the asymptotic monogamy/promiscuity entanglement behavior. In comparison, for strong couplings, for both the positive and negative coupling, we observe promiscuous behavior, not captured by the two-level approximation. 
\subsection{C. Low dimensional approximation of the QDO trimer Hamiltonian}
\begin{figure}
    \centering    \includegraphics{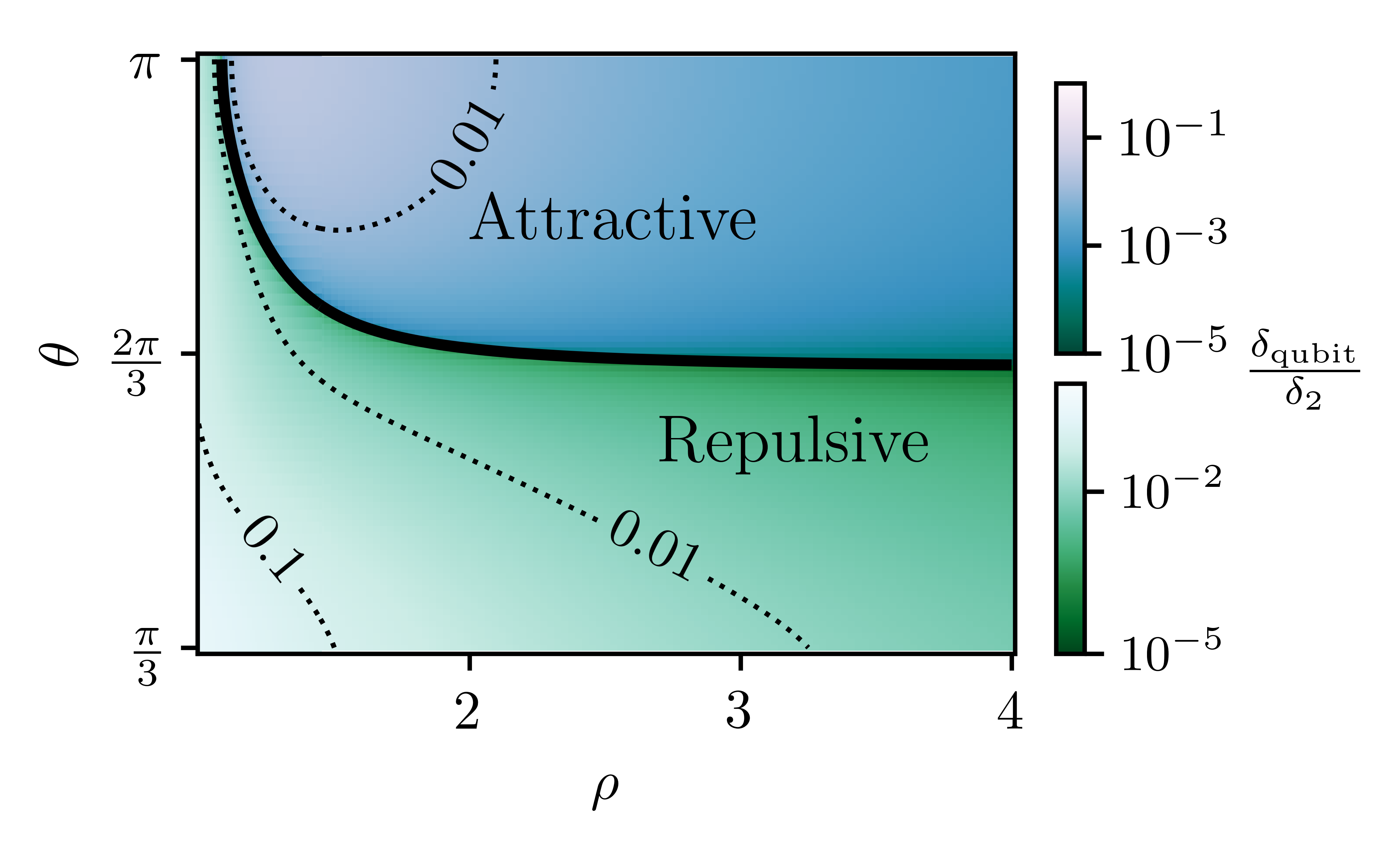}
    \caption{Behavior of  $\delta_{\mathrm{qubit}}$, normalized by the pairwise potential $\delta_{2}$ as a function of trimer parameters $\rho$ and $\theta$ (see main-text for trimer definition). The black line separates the phase space into regions where $\delta_{\mathrm{qubit}}>0$ and  
    $\delta_{\mathrm{qubit}}<0$.
    Both regions are shown as heat-maps, plotted on a log-scale. 
    }
    \label{isingtrimer}
\end{figure}
Here we use Eq. \eqref{qubit} to approximate the QDO Hamiltonian in the main-text, in Eq. (2). We set $d=3N$ and $\alpha \mathcal{T}_{ij}=w_{ij}$, in Eq. \eqref{qubit}, where $\mathcal{T}$ is the dipole-dipole interaction matrix and $\alpha$ is the QDO polarizability. The two level approximation of the QDO binding energy is given by,
\begin{equation}
    E_{\mathrm{qub}}=\frac{1}{2}\left(3N-\mathrm{min}[\mathrm{eig}(\hat{H}^{d=3N}_{\mathrm{qub}})]\right).
\end{equation}
We compare the approximate measure of binding given by the qubit Hamiltonian, with the approximate measure of binding given by the pairwise approximation in Eq. (4), in the main-text by defining,  $\delta_{\mathrm{qub}}=E_{\mathrm{qub}}-\delta_{2}$. This quantifies the extend to which Eq. \eqref{qubit} is able to capture the MB potential, for a $N$ QDO system.  

We  focus on the trimer setup, depicted in Fig 1 $a)$ in the main-text, which depends on geometry parameter $\theta$ and interactions strength $\rho$. For the trimer, the approximate qubit Hamiltonian in Eq. \eqref{qubit} is given by $\hat{H}^{d=9}_{\mathrm{qub}}$, as the trimer contains 9 modes. Fig. \ref{isingtrimer} shows the ability of the qubit approximation to capture the MB potential in the trimer. The sign of $\delta_{\mathrm{qub}}$ matches the behavior of the AT potential at weak couplings.  However by comparison with the heatmap in Fig 1 $c)$ in the main-text, Fig. \ref{isingtrimer} shows that the sign change in $\delta_{\rm MB}$ as a function of coupling strength is not captured by the two-level approximation. The two level qubit approximation is thus unable to explain the trimer physics at strong coupling.

\section{S9. Multipartite Tangle and Energy }
From the definition of the tangle $\tau_{G}(i)$ and the reduced tangle $\tilde{\tau}_{G}(i)$, both given in the main-text, we can write
\begin{equation}\label{tgtilde}
\tilde{\tau}_{G}=\frac{1}{4}\sum^{3N}_{i=1}\Delta_{i}=\sum^{3N}_{i=1}g(\tau_{G}(i)),
\end{equation}
where $\Delta_{i}=\langle \hat{\chi}^{2}_{i} \rangle\langle \hat{\mathcal{P}}^{2}_{i} \rangle-1$.
By defining $ \delta^{\chi}_{i}=\langle \hat{\chi}^{2}_{i} \rangle-1,\: \delta^{\mathcal{P}}_{i}=1-\langle \hat{\mathcal{P}}^{2}_{i} \rangle$, we write $\Delta_{i}$ as follows,
\begin{equation}\label{delta}
    \Delta_{i}=\delta^{\chi}_{i}-\delta^{\mathcal{P}}_{i}-\delta^{\chi}_{i}\delta^{\mathcal{P}}_{i}.
\end{equation}
By using the unitary invariance of the trace operator and writing $\left(\lambda_{i}\right)^{\pm 1/2}=\left(1+\alpha t_{i}\right)^{\pm 1/2}$, the binomial expansion of $\left(\lambda_{i}\right)^{\pm 1/2}$ in small parameter $\alpha t_{i}$ is
\begin{equation}\label{deltx}
\sum^{3N}_{i=1}\delta^{\chi}_{i}=\sum^{3N}_{i=1}\left(\left(\sqrt{\lambda_{i}}\right)^{-1}-1\right)=\sum^{\infty}_{k=2}{-0.5\choose k}\alpha^{k}\mathrm{tr}(\mathcal{T}^{k}),
\end{equation}
and 
\begin{equation}\label{deltp}
\sum^{3N}_{i=1}\delta^{\mathcal{P}}_{i}=\sum^{3N}_{i=1}\left(1-\sqrt{\lambda_{i}}\right)=-\sum^{\infty}_{k=2}{0.5\choose k}\alpha^{k}\mathrm{tr}(\mathcal{T}^{k}).
\end{equation}
By inserting Eq. \eqref{deltx} and Eq. \eqref{deltp} into Eq. \eqref{delta} and Eq. \eqref{tgtilde}, then simplifying the resultant expression we arrive at 
\begin{equation}\label{tangletilde}
    \tilde{\tau}_{G}=\sum^{\infty}_{k=2} (k-1)\delta_{k}-\frac{1}{4}  \sum^{3N}_{i=1}\delta^{\chi}_{i}\delta^{\mathcal{P}}_{i},
\end{equation}
where $\delta_{k}$ is defined in the main-text.
The upper bound on $\tilde{\tau}_{G}$ in Eq. 8, follows from  $\delta^{\chi}_{i}=\langle \hat{\chi}^{2}_{i} \rangle-1 \geq 0$ and  $\delta^{\mathcal{P}}_{i}=1-\langle \hat{\mathcal{P}}^{2}_{i} \rangle \geq 0$, both $\forall i \in \{1,..,3N\}$, due to a combination of lemma. 1 \ref{lem2} and lemma. 3 in S3.

\begin{figure}
    \centering
    \includegraphics{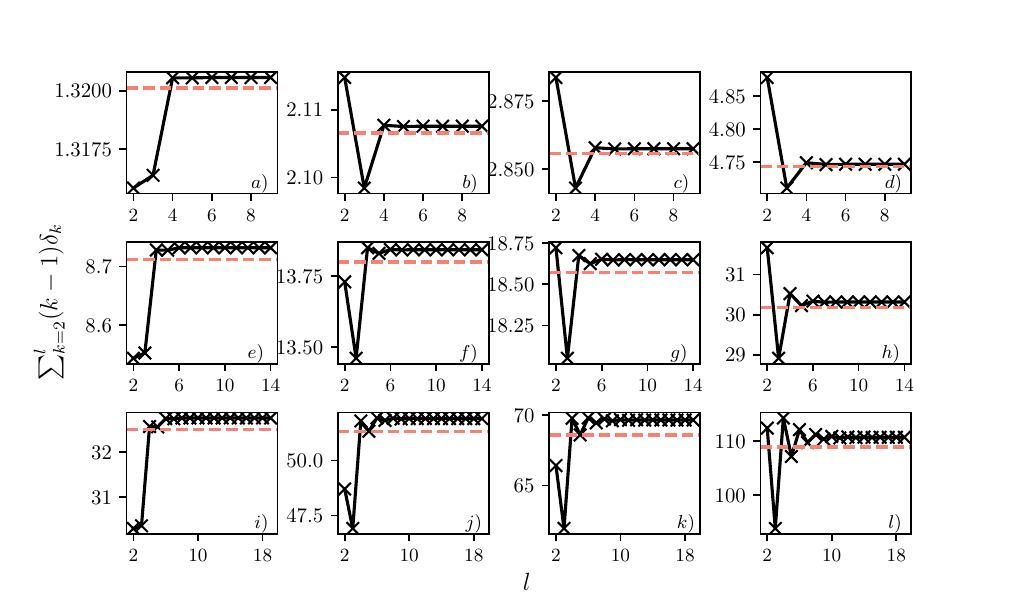}
    \caption{Convergence to Eq. 8, given in the main-text, for the honeycomb lattice in $a)$, $e)$ and $i)$, the square lattice in $b)$, $f)$ and $j)$ the triangular lattice in $c)$, $g)$ and $k)$ and the cubic lattice in $d)$, $h)$ and $l)$. Panels $a)-d)$, are at $\rho=4$, $e)-h)$ at $\rho=2.93$ and $i)-l)$ are at $\rho=2.37$. The dashed horizontal line shows the value of $\tilde{\tau}_{G}$ for the correspondent $\rho$ value and lattice.}
    \label{supflB}
\end{figure}

Given that $g(x)$ is a subadditive function i.e $g(x+y) \leq g(x)+g(y)$ (see S10),
\begin{equation}
    g\left(\tau_{G}=\sum^{3N}_{i=1}\tau_{G}(i)\right) \leq \sum^{3N}_{i=1}g(\tau_{G}(i))=\tilde{\tau}_{G}.
\end{equation}
This sub-additive property of $g(x)$ can be extend to the inequality in Eq. 8 in the main-text,
\begin{equation}\label{sec}
    g(\tau_{G}) \leq \tilde{\tau}_{G} \leq \sum^{\infty}_{k=2} (k-1)\delta_{k}.
\end{equation}
The sum of each of the $3N$ monogamy inequalities \cite{hiroshima2007monogamy} in the QDO ensemble are given by
\begin{equation}\label{totomong}
    \sum^{3N}_{i=1}\tau_{G}(i) \geq \sum^{3N}_{\substack{i,j=1\\ j \neq i}} \underbrace{f((-\langle \hat{\chi}_{i}\hat{\chi}_{j} \rangle \langle \hat{\mathcal{P}}_{i}\hat{\mathcal{P}}_{j}\rangle)}_{\tau^{\mathrm{sys}}_{G}(i:j)} \geq \\  \sum^{3N}_{\substack{i,j=1\\ j \neq i}}\tau_{G}(i:j),
\end{equation}
with $f(x)$ defined in the main-text. Inserting \eqref{totomong} into \eqref{sec}, we get
\begin{equation}
       g\left(\sum^{3N}_{\substack{i,j=1\\ j \neq i}}\tau^{\mathrm{sys}}_{G}(i:j) \right) \leq \sum^{\infty}_{k=2} (k-1)\delta_{k}.
\end{equation}

In Fig. \ref{supflB}, we show the convergence to the upper bound on the reduced tangle given in the main-text in Eq. 8, 
for the same four crystal lattices, investigated numerically in the main-text. We further consider the same setups (number of QDOs and open boundary conditions) as in the main-text.

\subsection{A. Convergence to the reduced tangle bounds on the cubic lattice}
\begin{figure}
    \centering
    \includegraphics{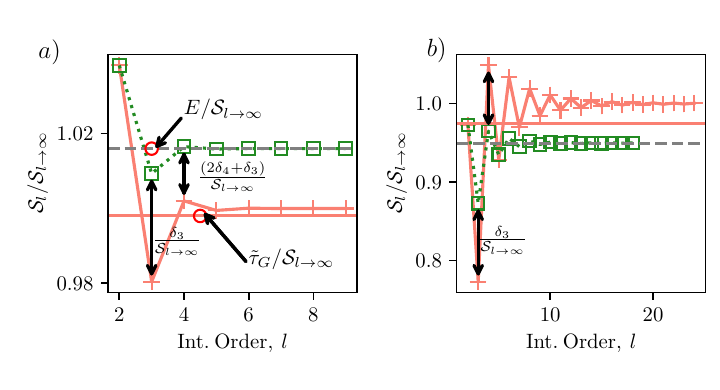}
    \caption{In $a)-b)$, where $\rho=3.4$ and $\rho=2.2$ respectively, the orange plus shaped markers show $S_{l}$, where $S_{l}=\sum^{l}_{k=1}(k-1)\delta_{k}$, normalized by $S_{l \rightarrow \infty}=\sum^{\infty}_{k=1}(k-1)\delta_{k}$ (further showing the convergence to Eq 8 in the main-text, in addition to Fig. \ref{supflB}) . This is shown for the $N=19 \times 19 \times 19$ cubic lattice. The horizontal line shows the normalized value of the reduced tangle, $\tilde{\tau}_{G}/\mathcal{S}_{l \rightarrow \infty}$. The green square markers show the series $\sum^{l}_{k=1}\delta_{k}$, also normalized by $\mathcal{S}_{l\rightarrow \infty}$. The grey dotted lines show the convergence of  $\sum^{l\rightarrow \infty}_{k=1}\delta_{k}$ to the dispersive bond energy $E$  for the two $\rho$ values.
    }
    \label{cubicbounds}
\end{figure}
In Fig. \ref{supflB}, we showed the convergence to the reduced tangle bound on the various lattice setups considered in the main-text. Here we will provide further analysis of the convergence in the cubic lattice. Having shown the tightness of the reduced tangle bound in the trimer system for Fig. 1 $d)-e)$, in the main-text, Fig. \ref{cubicbounds} show the tightness of the entanglement constraint at both strong and weak coupling in the cubic lattice. The orange plus shaped markers show $\mathcal{S}_{l}$ normalized by the value as $l \rightarrow \infty$. As defined in the Fig. \ref{cubicbounds} caption $\mathcal{S}_{l}$ is a short-hand notation for a finite approximation (up to order $l$) of the MB energy terms on the r.h.s on Eq. \eqref{tangletilde} (also the r.h.s of Eq 8 in the main-text).


In Fig. \ref{cubicbounds}, we show that, for the cubic lattice at $\rho=3.4$, $\delta_{2} >\tilde{\tau}_{G}$ but $\delta_{2} +2\delta_{3}<\tilde{\tau}_{G}$. This is the same behavior observed in the trimer at $\theta=\pi/2$, at large $\rho$ values. For stronger couplings in the cubic lattice, as shown by Fig. \ref{cubicbounds} $b)$, we note the slow convergence of $\mathcal{S}_{l}$ to $\mathcal{S}_{l\rightarrow \infty}$, where $\delta_{2}+2\delta_{3}+3\delta_{4}+4\delta_{5}+5\delta_{6}+6\delta_{7}<\tilde{\tau}_{G}$. Fig. \ref{cubicbounds} further shows the convergence of the series $\sum^{ \infty}_{k=1}\delta_{k}$ to the dispersive bond energy $E$, for the same $\rho$ values in the cubic lattice.
\section{S10. Properties of $g(x)$}
\textit{lemma. 4} $g(y) \geq g(x)$, given $y \geq x \geq 0$.

\textit{Proof.} The derivative of $g(x)$ is non-negative for any positive $x$,

\begin{equation}
    \frac{\partial\left(g(x)\right)}{\partial_{x}}=\frac{1+\sqrt{x}(1+2\sqrt{x}+2x)}{(1+2\sqrt{x})^{3}}\geq 0, \quad{} \forall x \geq 0.
\end{equation}

\textit{lemma. 5} $g(x+y) \leq g(x)+g(y)$, given $x,y \geq 0$

\textit{Proof.} The derivative of $g(x)/x$ is always negative for any positive $x$,
\begin{equation}
    \frac{\partial\left(\frac{g(x)}{x}\right)}{\partial_{x}}=\frac{-(1+\sqrt{x})}{(1+2\sqrt{x})^{3}\sqrt{x}} \leq 0, \quad{} \mathrm{given} \quad{} x \geq 0.
\end{equation}
As a result for positive $x$ and $y$, we can write,
\begin{equation}\label{ineqg1}
    \frac{g(y)}{y} \leq \frac{g(x)}{x}, \quad{} \mathrm{if} \quad{} 0 \leq x \leq y.
\end{equation}
Thus,
\begin{equation}\label{ineqg2}
    g(cx) \leq cg(x) \quad{} c \geq 1,
\end{equation}
because $c$ can always be written as $c=y/x$. As a result the inequality in Eq. \eqref{ineqg2} follows from Eq. \eqref{ineqg1}. The subadditivity of $g(x)$ follows from inserting $c_{1}=1+y/x$ into $c$ in Eq. \eqref{ineqg2} and $c_{2}=1+x/y$ into $c$ in Eq. \eqref{ineqg2} also. By adding and simplifying the resultant two inequalities, Eq. \eqref{ineqg2} implies $g(x+y) \leq g(x)+g(y)$, for positive $x$ and $y$.

\begin{figure}
    \centering
    \includegraphics{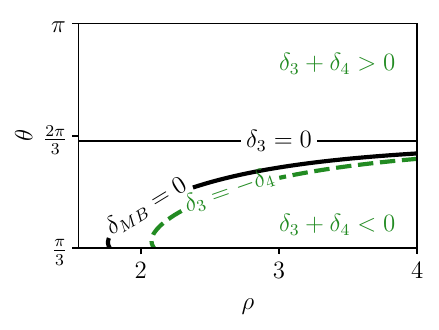}
    \caption{The green dotted curve, shows the $\rho$ and $\theta$ values where $\delta_{3}=-\delta_{4}$. The black solid lines are included for comparison, showing where the MB effects are zero $\delta_{MB}=0$ and where the AT potential is zero, $\delta_{3}=0$.}
    \label{delta4trimer}
\end{figure}
\begin{figure}
    \centering
    \includegraphics{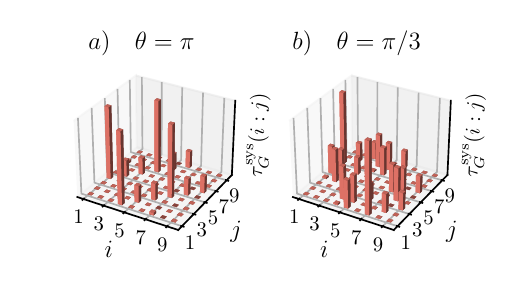}
    \caption{$a)-b)$ show the maximal allowed pairwise tangle, for each pair of modes in the trimer ground state, consistent with the associated monogamy inequalities, given in Eq. \eqref{totomong}. $a)$ shows the linear three QDO arrangement and  $b)$ shows the triangular geometry. Both $a)-b)$ are shown for $\rho=2.60$. }
    \label{barchartpairwise}
\end{figure}

\section{S11. Approximating the Many-body potential in the trimer}
Here we consider the trimer system, defined in the main-text. In Fig. \ref{delta4trimer}, we show the qualitative agreement between boundary separating the repulsive and attractive MB effects and dotted green curve, separating the region of phase space between where $|\delta_{3}|<|\delta_{4}|$  and where $|\delta_{3}|>|\delta_{4}|$. This is shown in the main-text for the various lattices in Fig 2, by where the large orange dotted curve passes through zero and where the solid grey curve passes through zero, for the respective lattices. 

Fig. \ref{delta4trimer}, further shows that for a given $\rho$ value, the deviation in $\theta$ value between where  $\delta_{3}=-\delta_{4}$ and where $\delta_{MB}=0$, grows for the more acute the trimer geometries. This is also observed for the various extended lattices where the approximation $\delta_{3}+\delta_{4} \approx \delta_{MB}$, works best for the Honeycomb lattice, which has an obtuse three-body geometry. 

\section{S12. Distribution of $\tau^{\mathrm{sys}}_{G}(i:j)$ in the trimer}
Upper bounds on the sum of the pairwise tangle in the QDO ground state is given in Eq. \eqref{totomong}.
See Fig. \ref{barchartpairwise}, for a measure of the strength of the quantum correlations, for each pair of modes in the 9 mode trimer system, for different geometries. Note that there is a greater number of non-trivial values of $\tau^{\mathrm{sys}}_{G}(i:j)$ in the triangular configuration, compared with the linear arrangement. 

\section{S13. MB potential and the entanglement distribution in the linear to zigzag chain}

\begin{figure}
    \centering
    \includegraphics{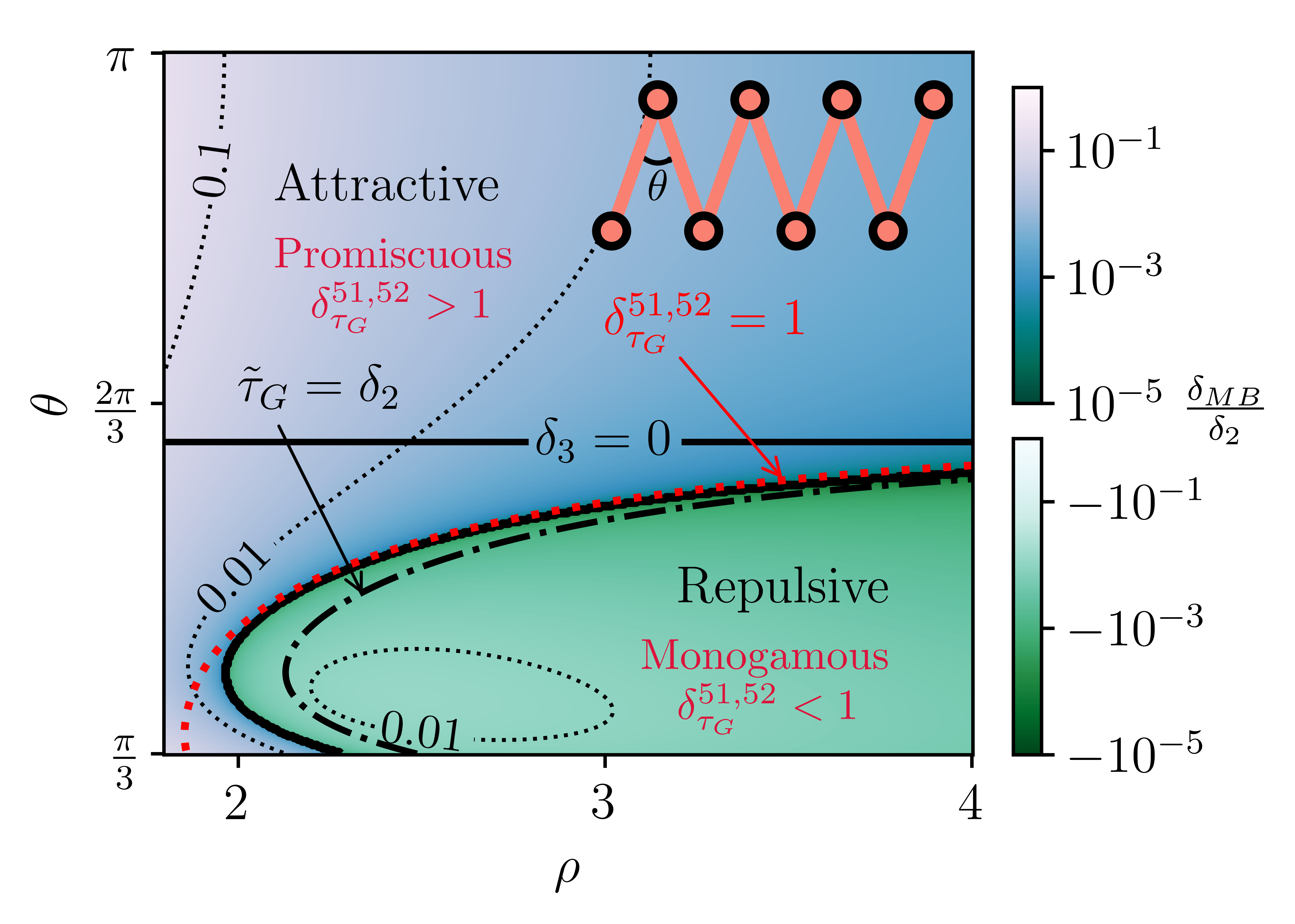}
    \caption{ Above the red dotted line the sum of the pairwise tangle shared between the nearest neighbour QDOs at the center of the chain behaves promiscuously, whereas below they behave monogamously. For $\rho \geq 2$, the monogamy/promiscuity boundary overlaps the black line where MB energy corrections vanish and the pairwise additive approximation holds; above which MB corrections are attractive and below which are repulsive; both shown as heat-maps, plotted on a log scale. }
    \label{chainheatmap}
\end{figure}

In the main-text we established the close alignment between the monogamous or promiscuous behavior of the entanglement distribution and the sign of the MB effects in the trimer. Here we extend this analysis to a chain of QDOs.  The chain configuration is depicted in the inset of Fig. \ref{chainheatmap}. Given that $R_{\mu,\mu+1}=R$, the coordinates of the QDOs are as follows: $(0,0,0)$,  $(R\sin{[\theta/2]},R\cos{[\theta/2]},0)$, $(2R\sin{[\theta/2]},0,0)$, $(3R\sin{[\theta/2]}),R\cos{[\theta/2]},0)$, $(4R\sin{[\theta/2]},0,0)$, 
$...$ 
$,((N+1)R\sin{[\theta/2]}$
$,R\cos{[\theta/2]},0)$. The inset in Fig. \ref{chainheatmap}, depicts the chain in it's zigzag configuration, where $\theta=\pi/3$.  Here we focus on the MB effects and entanglement properties of the chain, going from the linear to zigzag configuration, with chain geometry determined by $\theta$. Fig. \ref{chainheatmap} shows a heat-map of $\delta_{MB}/\delta_{2}$, as a function of $\rho$ and $\theta$ in the linear-zigzag chain, where we set $N=100$. We consider 400 $\theta$ values in the range $
\theta \in \{\pi ,\pi/3\}$ and 400 $\rho$ values in the range $\rho \in \{4,1.8\}$. The chain has open boundary conditions.

We define the promiscuous or monogamous entanglement distribution
in the quasi- 1D chain, based on whether the net maximal (monogamy allowed) pairwise
tangle shareable between two QDOs close to the center of the chain are greater than ($\delta^{N/2+1,N/2+2}_{\tau_{G}}>1$) or less than ($\delta^{N/2+1,N/2+2}_{\tau_{G}}<1$),  the net reference pairwise tangle. Similarly to Fig. \ref{chainheatmap}, Fig. \ref{chainheatmap} shows how the parameter space of the linear-zigzag chain is partitioned into regions, separated by the dotted red line, where  $\delta^{N/2+1,N/2+2}_{\tau_{G}}$ is either enhanced (above) or suppressed (below).  The red dotted line, partitioning the parameter space in promiscuous and monogamous entanglement, as a function of $\rho$ and $\theta$ is given by $\delta^{51,52}_{\tau_{G}}=1$, in the 100 QDO chain. For $\rho \geq 2$, this red dotted line closely overlaps where $\delta_{MB}=0$. At $\rho =2.3$, the deviation in $\theta$ value between the two boundaries is less than 1\% of the $\theta$ value where the MB effects vanish. Thus away from strong couplings, we thus see how the behavior of entanglement in a central pair of QDOs in the chain, well captures the sign of the MB potential.

\end{document}